\documentclass[aps,prc,10pt,showpacs,showkeys,twocolumn,superscriptaddress,groupedaddress]{revtex4-1}

\usepackage[english]{babel}
\usepackage{indentfirst}
\usepackage{amssymb}
\usepackage{braket}
\usepackage{bm}
\usepackage{amsxtra}
\usepackage{amsmath}
\usepackage{supertabular}
\usepackage{multirow}
\usepackage{color}
\usepackage [mathcal]{eucal}
\usepackage{graphicx}
\usepackage{graphics}
\usepackage{comment}

\def\dblone{\hbox{$1\hskip -1.2pt\vrule depth 0pt height 1.6ex width 0.7pt
     \vrule depth 0pt height 0.3pt width 0.12em$}}
\def\n3lo{$\mathrm{N}^3\mathrm{LO}$}
\def\chiral4lo{$\mathrm{N}^4\mathrm{LO}$}
\def\3bn2lo{$\mathrm{N}^2\mathrm{LO}$}

\providecommand{\csletcs}[2]{%
  \expandafter\let\csname#1\expandafter\endcsname\csname#2\endcsname}
\newcommand{\equivUC}[3][\def]{%
  \expandafter#1\csname#2\expandafter\endcsname\expandafter{%
    \csname#3\endcsname}}
\newcommand{\UPCASElang}[1]{
  \uppercase{\def\temp{#1}}
  \csletcs{l@\temp}{l@#1}
  \equivUC{date\temp}{date#1}
  \equivUC{captions\temp}{captions#1}
  \equivUC{extras\temp}{extras#1}
  \equivUC{noextras\temp}{noextras#1}
  \equivUC[\edef]{\temp hyphenmins}{#1hyphenmins}
}
\UPCASElang{english}
\UPCASElang{italian}

\begin{document}

\noindent
\title{Theoretical Optical Potential Derived From Nucleon-Nucleon Chiral Potentials}

\author{Matteo Vorabbi$^{1}$}
\author{Paolo Finelli$^{2}$}
\author{Carlotta Giusti$^{1}$}
\affiliation{$~^{1}$Dipartimento di Fisica,  
Universit\`a degli Studi di Pavia and \\
INFN, Sezione di Pavia,  Via A. Bassi 6, I-27100 Pavia, Italy}

\affiliation{$~^{2}$Dipartimento di Fisica e Astronomia, 
Universit\`{a} degli Studi di Bologna and \\
INFN,
Sezione di Bologna, Via Irnerio 46, I-40126 Bologna, Italy}

\date{\today}

\begin{abstract} 

{\bf Background:} Elastic scattering is probably the main event in the interactions of nucleons with nuclei. Even if this process has 
been extensively studied in the last years, a consistent description, i.e. starting from microscopic two- and many-body forces connected by the same symmetries and principles, is still under development.
{\bf Purpose:} In this work we study the domain of applicability of microscopic two-body chiral potentials in the construction of an optical potential. 
{\bf Methods:} We basically follow the KMT approach \cite{Kerman1959551} to build a microscopic complex optical potential and 
then we perform some test calculations on $^{16}$O at different energies.
{\bf Results:}. Our conclusion is that a particular set of potentials with a Lippmann-Schwinger cutoff
at relatively high energies (above $500$ MeV)
has the best performances reproducing the scattering observables.
{\bf Conclusions:} Our work shows that building an optical potential within Chiral Perturbation Theory is a promising approach to the description of 
elastic proton scattering, in particular, in view of the future inclusion of many-body forces that naturally arise in such framework.

\end{abstract}


\maketitle


\maketitle

\section{Introduction}
Elastic proton scattering has been extensively studied over
many decades, both experimentally and theoretically, and there now exist extensive measurements of cross sections and polarization
observables for the elastic scattering of protons on a wide variety of stable nuclei in a wide range of energies.
A suitable and successful framework to describe the nucleon-nucleus ($NA$) interaction in the elastic scattering is provided by the
nuclear optical potential \cite{hodgson1963}.
With this instrument we can compute the scattering observables, such as the differential cross section, the analyzing power, and the
spin rotation, for the elastic $NA$ scattering across wide regions of the nuclear landscape.
The use of the optical potential has been extended to calculations of inelastic scattering and to generate the distorted waves for the
analysis of the cross sections for a wide variety of nuclear reactions. For instance, in quasi-elastic electron scattering an optical
potential is commonly used to describe the final-state interaction between the emitted nucleon and the residual nucleus in the exclusive
$(e,e^{\prime}p)$ \cite{Boffi1996} and in the inclusive $(e,e^{\prime})$ reactions \cite{capuzzi1991,meucci2003}.

The optical potential can be obtained in different ways. It can be obtained phenomenologically \cite{Varner:1991zz,Koning:2003zz}, by assuming a form of the potential
and a dependence on a number of adjustable parameters for the real and imaginary parts that characterize the shape of the nuclear
density distribution and that vary with the nucleon energy and with the nucleus mass number. These parameters are adjusted to
optimize the fit to the experimental data of elastic $NA$ scattering. The optical potential has an imaginary part that takes into
account the absorption of the reaction flux from the elastic channel to the non-elastic reaction channels. Alternatively and more
fundamentally, the optical potential can be obtained microscopically. The calculation requires, in principle, the solution of the full
nuclear many-body problem, which is beyond present capabilities. In practice, some approximations must necessarily be adopted.
With suitable approximations, microscopic optical potentials are usually derived from two basic quantities: the nucleon-nucleon ($NN$)
$t$ matrix and the matter distribution of the nucleus. All these models based on the $NN$ interaction are nonrelativistic (see 
Ref.~\cite{negele2006advances} for a detailed review).
Because microscopic optical potentials do not
contain adjustable parameters, we expect that they have a greater predictive power when
applied to situations where experimental data are not yet available, such as, for instance, to the study of unstable nuclei.

The theoretical justification for the $NA$ optical potential built in terms of underlying $NN$ scattering amplitudes was given
for the first time by Chew \cite{PhysRev.80.196} and Watson {\it et al.} \cite{PhysRev.89.575,PhysRev.92.291} more than $60$ years ago.
Successively, Kerman, McManus, and Thaler (KMT) \cite{Kerman1959551} developed the Watson multiple scattering approach expressing the
$NA$ optical potential by a series expansion in terms of the free $NN$ scattering amplitudes. Several years later, with the development
of high accuracy $NN$ potentials, there has been a renewed interest in finding a rigorous treatment of the $NA$ scattering theory
in momentum space. Such potentials permit to generate the $NN$ interaction directly in momentum space, that is thus chosen as the
working space in which to develop the $NA$ optical potential and to compute the elastic scattering observables. Several authors
contributed to the development of the multiple scattering theory and, with a series of papers \cite{PhysRevC.41.2257,PhysRevC.46.279,
PhysRevC.48.351,PhysRevC.50.2995,PhysRevLett.63.605,PhysRevC.41.2188,PhysRevC.42.652,PhysRevC.43.1875,PhysRevC.50.2480,PhysRevC.52.301,
PhysRevC.54.2570,PhysRevC.84.034606,PhysRevC.40.881,PhysRevC.41.814,PhysRevC.44.1569,elster1993,PhysRevC.47.2242,PhysRevC.48.2956,
PhysRevC.52.1992,PhysRevC.56.2080,orazbayev2013}, to calculations of microscopic optical potentials. The present work is framed in
this context.

The $NN$ potential is an essential ingredient in the $NA$ scattering theory and its off-shell properties play an important role.
To obtain a good description of these properties, the optical potential models have always employed 
``realistic''  potentials, in which  the experimental $NN$ phase shifts are reproduced with a $\chi^2$ per data $\simeq 1$. The most commonly used $NN$ potentials are those given by groups from Nijmegen \cite{PhysRevC.49.2950},
Paris \cite{PhysRevC.21.861}, Bonn \cite{machleidt1989}, Argonne \cite{PhysRevC.51.38}.
In contrast, with recent advances in Lattice Quantum Chromodynamics (Lattice QCD), new attempts have been made to derive the nuclear potential directly
from first principles \cite{Aoki:2012tk}.
However, since QCD is a non-perturbative theory in the low-energy regime, that is characteristic of nuclear physics, an {\it ab-initio}
solution of this problem is not feasible at the moment. On the other hand, when the concepts of Effective Field Theory (EFT) were applied
to low-energy QCD, Chiral Perturbation Theory (ChPT) \cite{Scherer:2002tk} 
was developed and it became possible to implement chiral symmetry
consistently in a theory of pionic and nuclear interactions. For this theory some sort of perturbative expansion was assumed,
such that only a finite number of terms contribute at a given order. This expansion was provided by powers of small external momenta
over the chiral symmetry breaking scale, $\Lambda_{\chi} \approx 1$ GeV. The $NN$ potential is then calculated perturbatively in the
chiral expansion and iterated to all orders in a Schr\"odinger or Lippmann-Schwinger (LS) equation to obtain the nuclear amplitude.

The most recent available chiral potentials are developed at fourth order (\n3lo) in the chiral expansion
and are used in this work as a
basic ingredient to compute the $NN$ $t$ matrix for the construction of the $NA$ optical potential. In particular, in all the
calculations presented in this paper we adopt the two different versions of chiral potentials developed by Entem and
Machleidt (EM, Refs. \cite{chiralmachleidt,PhysRevC.88.054002,PhysRevC.87.014322,PhysRevC.91.054311,PhysRevC.75.024311}), and Epelbaum, Gl\"ockle, and Mei\ss ner (EGM, Ref. \cite{chiralepelbaum}). 

Very recently Epelbaum {\it et al.} \cite{PhysRevLett.115.122301} presented a nucleon-nucleon potential at fifth order (\chiral4lo)
with an innovative coordinate-space regularization. 
It is also worth to mention that in the last years some authors  \cite{PhysRevC.88.024614,PhysRevC.90.051601,0954-3899-42-2-025104}, following different approaches, have started to include 
chiral three-body forces \cite{Machleidt20111, RevModPhys.81.1773} at order \3bn2lo.  
We plan to extend our calculations along these research lines in forthcoming papers.

The second important ingredient of the $NA$ scattering theory is the microscopic structure of the nuclear target, given by neutron
and proton densities. These quantities are computed within the Relativistic Mean-Field (RMF) description \cite{Niksic20141808} of
spherical nuclei using a Density-Dependent Meson-Exchange (DDME) model, where the couplings between mesonic and baryonic fields are
assumed as functions of the density itself \cite{PhysRevC.66.024306}.

The paper is organized as follows.
In Section~\ref{theofram} we describe the theoretical framework used to calculate the $NA$ optical potential and
the scattering observables. This section is divided into five subsections in which we outline the different aspects of the calculation.
In Subsection~\ref{firstop} the general scattering problem is introduced in the momentum frame and it is represented by the LS equation
for the entire system composed by the projectile and the target nucleus. This equation is separated into a simple one-body equation for
the transition matrix and a more complicated one for the optical potential. Using the optimum factorization approximation the expression
for the optical potential is then reduced to a simple form, in which the $NN$ $t$ matrix and the nuclear density are factorized.
In Subsection~\ref{nnamp} we give the explicit formulae to compute the $NN$ Wolfenstein amplitudes that are proportional to the central
and spin-orbit parts of the $NN$ $t$ matrix, that is then used to compute the $NA$ optical potential.
In Subsections~\ref{transparwave} and \ref{scatobs} we describe the theoretical framework to solve the $NA$ LS equation in the partial
wave representation and then we use these solutions to compute the scattering observables. In Subsection~\ref{coulpot} we show the
algorithm we use to include in the model the Coulomb interaction between the projectile and the target nucleus.

In Section~\ref{nnresults} we present and discuss the theoretical results for the $NN$ Wolfenstein amplitudes obtained with the
different $NN$ potentials. In particular, the novelty in our calculations is the use of the chiral
potentials \cite{chiralmachleidt,chiralepelbaum} as basic ingredient to compute the microscopic $NA$ optical potential and
the scattering observables.

In Section~\ref{scattresults} we present theoretical results for the scattering observables on $^{16}$O calculated with all
$NN$ potentials. Predictions based on EM and EGM potentials are 
compared with available experimental data in order to determine the most successful theoretical approach and 
the best LS cutoff.

Finally, in Section~\ref{concl} we draw our conclusions.

\section{Theoretical Framework}
\label{theofram}
\subsection{The First-Order Optical Potential}
\label{firstop}
The general problem of the elastic scattering of a proton from a target nucleus of $A$ nucleons can be stated in momentum space by
the full $(A+1)$-body Lippmann-Schwinger equation
\begin{equation}\label{generalscatteq}
T = V + V G_0 (E) T \, ,
\end{equation}
whose general solution is beyond present capabilities. A reliable method to treat Eq.~(\ref{generalscatteq}) is given by the
spectator expansion \cite{PhysRevC.52.1992}, in which the multiple scattering theory is expanded in a finite series of terms
where the target nucleons interact directly with the incident proton. In particular, the first term of this series only involves
the interaction of the projectile with a single target nucleon, the second term involves the interaction of the projectile with two
target nucleons, and so on to the subsequent orders. In the standard approach to elastic scattering, Eq.~(\ref{generalscatteq})
is separated into two equations. The first one is an integral equation for $T$
\begin{equation}\label{firsttamp}
T = U + U G_0 (E) P T \, ,
\end{equation}
where $U$ is the optical potential operator, and the second one is an integral equation for $U$
\begin{equation}\label{optpoteq}
U = V + V G_0 (E) Q U \, .
\end{equation}
The operator $V$ represents the external interaction and the total Hamiltonian for the $(A+1)$-nucleon system
is given by
\begin{equation}
H_{A+1} = H_0 + V \, .
\end{equation}
If we assume the presence of only two-body forces, the operator $V$ is expressed as
\begin{equation}
V = \sum_{i=1}^A v_{0i} \, ,
\end{equation}
where the two-body potential $v_{0i}$ describes the interaction between the incident proton and the {\it i}th target nucleon.
The system is asymptotically an eigenstate of the free Hamiltonian $H_0$ and $G_0 (E)$ is the free propagator for the $(A+1)$-nucleon
system
\begin{equation}
G_0 (E) = \frac{1}{E - H_0 + i \epsilon} \, .
\end{equation}
The free Hamiltonian is given by
\begin{equation}
H_0 = h_0 + H_A \, ,
\end{equation}
where $h_0$ is the kinetic energy operator of the projectile and $H_A$ is the target Hamiltonian,
\begin{equation}
H_A \ket{\Phi_A} = E_A \ket{\Phi_A} \, ,
\end{equation}
where $\ket{\Phi_A}$ is the ground state of the target.
The operators $P$ and $Q$ in Eqs.~(\ref{firsttamp}) and (\ref{optpoteq}) are projection operators,
\begin{equation}
P + Q = \dblone \, ,
\end{equation}
and $P$ fulfills the condition
\begin{equation}\label{procommutator}
[G_0 , P] = 0 \, .
\end{equation}
In the case of elastic scattering $P$ projects onto the elastic channel and can be defined as
\begin{equation}
P = \frac{\ket{\Phi_A} \bra{\Phi_A}}{\braket{\Phi_A|\Phi_A}} \, .
\end{equation}
With these definitions, the elastic transition operator may be defined as $T_{\mathrm{el}} = PTP$, and, in this
case, Eq.~(\ref{firsttamp}) becomes
\begin{equation}\label{elastictransition}
T_{\mathrm{el}} = P U P + P U P G_0 (E) T_{\mathrm{el}} \, .
\end{equation}
Thus the transition operator for elastic scattering is given by a one-body integral equation. In order to solve
Eq.~(\ref{elastictransition}) we need to know the operator $P U P$. In the spectator expansion the operator $U$ is expanded as
\begin{equation}\label{spectatorexp}
U = \sum_{i=1}^A \tau_i + \sum_{i,j\neq i}^A \tau_{ij} + \sum_{i,j\neq i,k\neq i,j}^A \tau_{ijk} + \cdots \, ,
\end{equation}
according to the number of nucleons interacting with the projectile. In the present work we only consider the
first-order term of this expansion and thus the optical potential operator becomes
\begin{equation}
U = \sum_{i=1}^A \tau_i \, ,
\end{equation}
where $\tau_i$ can be expressed as
\begin{equation}\label{hattaueq}
\tau_i = \hat{\tau}_i - \hat{\tau}_i G_0 (E) P \tau_i \, .
\end{equation}
For elastic scattering we only need to consider $P \tau_i P$, or, equivalently,
\begin{equation}\label{hattaueq2}
\begin{split}
\braket{\Phi_A |\tau_i |\Phi_A} &= \braket{\Phi_A |\hat{\tau}_i |\Phi_A} - \braket{\Phi_A |\hat{\tau}_i |\Phi_A} \\
&\times \frac{1}{(E-E_A) - h_0 + i \epsilon} \braket{\Phi_A |\tau_i |\Phi_A} \, ,
\end{split}
\end{equation}
where $\hat{\tau}_i$ is the solution of
\begin{equation}\label{hattaueq3}
\hat{\tau}_i = v_{0i} + v_{0i} G_0 (E) \hat{\tau}_i \, .
\end{equation}
Expanding the propagator $G_0 (E)$ within a single-particle description, at the first order we obtain
\begin{equation}
G_i (E) = \frac{1}{(E-E^i) - h_0 - h_i - W_i + i \epsilon} \, ,
\end{equation}
where $h_i$ is the kinetic energy of the {\it i}th target nucleon and $W_i$ is given by
\begin{equation}
W_i = \sum_{j \neq i} v_{ij} \, ,
\end{equation}
and represents the force between the struck nucleon and the other $(A-1)$ nucleons. 
With the operator $G_i (E)$, the $\hat{\tau}_i$ matrix of Eq.~(\ref{hattaueq3}) is expressed as
\begin{equation}\label{tildetaueq}
\hat{\tau}_i = v_{0i} + v_{0i} G_i (E) \hat{\tau}_i = t_{0i} + t_{0i} g_i W_i G_i (E) \hat{\tau}_i \, ,
\end{equation}
where the operators $t_{0i}$ and $g_i$ are defined as
\begin{equation}\label{freetmatrix}
t_{0i} = v_{0i} + v_{0i} g_i t_{0i} \, ,
\end{equation}
\begin{equation}
g_i = \frac{1}{(E-E^i) - h_0 - h_i + i \epsilon} \, .
\end{equation}
In Eq.~\ref{freetmatrix} the matrix $t_{0i}$ represents the free $NN$ $t$ matrix and in the Impulse Approximation (IA) we have
$\hat{\tau}_i \approx t_{0i}$. Thus in this approximation we only have to solve a two-body equation.

In order to develop a theoretical framework to compute the optical potential and the transition amplitude for the elastic
scattering observables, we follow the path outlined in Ref.~\cite{PhysRevC.30.1861}, that is based on the KMT multiple scattering
theory and that, at the first order, is equivalent to the IA. In this formulation, the elastic scattering amplitude
is given by
\begin{equation}
T_{\mathrm{el}} ({\bm k}^{\prime},{\bm k};E) = \frac{A}{A-1} \hat{T} ({\bm k}^{\prime},{\bm k};E) \, ,
\end{equation}
where the auxiliary elastic amplitude is determined by the solution of the integral equation
\begin{equation}\label{lipschw}
\hat{T} ({\bm k}^{\prime},{\bm k};E) = \hat{U} ({\bm k}^{\prime},{\bm k};\omega)
+ \int d^3 p \frac{\hat{U} ({\bm k}^{\prime},{\bm p};\omega) \,
\hat{T} ({\bm p},{\bm k};E)}{E (k_0) - E (p) + i \epsilon} \, ,
\end{equation}
and the auxiliary first-order optical potential is defined by
\begin{equation}\label{optpot}
\hat{U} ({\bm k}^{\prime},{\bm k};\omega) = (A-1) \braket{{\bm k}^{\prime} , \Phi_A | t (\omega) |{\bm k} , \Phi_A} \, ,
\end{equation}
where $t$ is any one of the free $NN$ $t_{0i}$ matrices. Our problem is then described in the zero-momentum frame of the $NA$ system
by Eq.~(\ref{lipschw}), where $k_0$ is the initial on-shell momentum and $E (k_0)$ is the corresponding initial energy of the system
in the $NA$ frame. To compute the scattering observables we only need the on-shell term $T_{\mathrm{el}} ({\bm k}_0 , {\bm k}_0;E)$ of
the transition matrix, but in this work we consider the full off-shell matrix with the general initial and final momenta ${\bm k}$
and ${\bm k}^{\prime}$, respectively.

The KMT first-order optical potential is given by Eq.~(\ref{optpot}) where $t (\omega)$ is
the free $NN$ $t$ matrix evaluated at a fixed energy $\omega$. Defining the new variables
\begin{equation}\label{qKvariables}
{\bm q} \equiv {\bm k}^{\prime} - {\bm k} \, , \qquad {\bm K} \equiv \frac{1}{2} ({\bm k}^{\prime} + {\bm k}) \, ,
\end{equation}
some manipulations \cite{PhysRevC.30.1861} give Eq.~\ref{optpot} in a factorized form (optimum factorization approximation) as the
product of the $NN$ $t$ matrix and the nuclear matter density,
\begin{equation}\label{optimumfact}
\begin{split}
\hat{U} ({\bm q},{\bm K};\omega) &= \frac{A-1}{A} \, \eta ({\bm q},{\bm K}) \\
&\times \sum_{N = n,p} t_{pN} \left[{\bm q},\frac{A+1}{A} {\bm K} ; \omega \right] \, \rho_N (q) \, ,
\end{split}
\end{equation}
where $N=n,p$, $t_{pN}$ represents the proton-proton ({\it pp}) and proton-neutron ({\it pn}) $t$ matrix, $\rho_N$ the neutron and proton profile
density, and $\eta ({\bm q},{\bm K})$ is the M\o ller factor,
\begin{equation}
\begin{split}
&\eta ({\bm q},{\bm K}) = \\
&{\left[ \frac{E_{\mathrm{proj}} ({\bm \kappa}^{\prime}) \, E_{\mathrm{proj}}
(-{\bm \kappa}^{\prime}) \, E_{\mathrm{proj}} ({\bm \kappa}) \, E_{\mathrm{proj}} (-{\bm \kappa})}{E_{\mathrm{proj}}
({\bm k}^{\prime}) \, E_{\mathrm{proj}} \left(- \frac{{\bm q}}{2} - \frac{{\bm K}}{A} \right) \,
E_{\mathrm{proj}} ({\bm k}) \, E_{\mathrm{proj}} \left(\frac{{\bm q}}{2} - \frac{{\bm K}}{A} \right)}
\right]}^{\frac{1}{2}} \, ,
\end{split}
\end{equation}
that imposes the Lorentz invariance of the flux when we pass from the $NA$ to the $NN$ frame in which the $t$ matrices are evaluated.
The optical potential obtained so far is an operator in the spin space of the projectile. To make the spin dependence explicit,
the $t$ matrix $t_{pN}$ is averaged over the spin of the struck nucleon and is written as
\begin{equation}\label{tmatdecomp}
\begin{split}
t_{pN} &\left[ {\bm q},\frac{A+1}{A} {\bm K};\omega \right] = t_{pN}^c \left[ {\bm q}, \frac{A+1}{A} {\bm K};\omega \right] \\
&+ \left( \frac{A+1}{2 A} \right) \frac{i}{2} {\bm \sigma}
\cdot {\bm q} \times {\bm K} \, t_{pN}^{ls} \left[ {\bm q}, \frac{A+1}{A} {\bm K};\omega \right] \, .
\end{split}
\end{equation}
The first term of Eq.~(\ref{tmatdecomp}) corresponds to the central spin-independent contribution and the second term corresponds
to the spin-orbit contribution. In the latter term the usual total Pauli spin operator of the $NN$ system is replaced by the Pauli spin
operator of the projectile, because the spin operator of the struck nucleon has been eliminated by the trace over the spin.
The replacement of Eq.~(\ref{tmatdecomp}) into Eq.~(\ref{optimumfact}) gives the optical potential as
\begin{equation}\label{opticaldec}
\hat{U} ({\bm q},{\bm K};\omega) = \hat{U}^c ({\bm q},{\bm K};\omega) + \frac{i}{2} {\bm \sigma} \cdot {\bm q} \times {\bm K} \,
\hat{U}^{ls} ({\bm q},{\bm K};\omega) \, ,
\end{equation}
where the central and the spin-orbit terms are given by
\begin{equation}\label{opticalcentral}
\begin{split}
\hat{U}^c ({\bm q},{\bm K};\omega) &= \frac{A-1}{A} \, \eta ({\bm q},{\bm K}) \\
&\times \sum_{N = n,p} t_{pN}^c \left[{\bm q},\frac{A+1}{A} {\bm K} ; \omega \right] \, \rho_N (q) \, ,
\end{split}
\end{equation}
\begin{equation}
\begin{split}
\hat{U}^{ls} ({\bm q},{\bm K};\omega) &= \frac{A-1}{A} \, \eta ({\bm q},{\bm K}) \left( \frac{A+1}{2 A} \right) \\
&\times \sum_{N = n,p} t_{pN}^{ls} \left[{\bm q},\frac{A+1}{A} {\bm K} ; \omega \right] \, \rho_N (q) \, .
\end{split}
\end{equation}
The optimally factorized optical potential given in Eqs.~(\ref{opticaldec}) and (\ref{opticalcentral}) exhibits nonlocality and
off-shell effects through the dependence of $\eta$ and $t_{pN}$ upon ${\bm K}$. The energy $\omega$ at which the matrices
$t_{pN}^c$ and $t_{pN}^{ls}$ are evaluated is fixed as
\begin{equation}
\omega = \frac{T_{lab}}{2} = \frac{1}{2} \frac{k_{lab}^2}{2 m} \, ,
\end{equation}
where $k_{lab}$ is the on-shell momentum of the projectile in the laboratory system. This is the fixed beam energy approximation, which
is a hystoric choice performed in all calculations based on the KMT formulation. A review of this type of calculations can be
found in Ref.~\cite{Ray1992223}.

\subsection{The {\it NN} Transition Matrix}
\label{nnamp}
The $NN$ elastic scattering amplitude for the scattering from a relative momentum ${\bm \kappa}$ to ${\bm \kappa}^{\prime}$, denoted
with $M ({\bm \kappa}^{\prime},{\bm \kappa},\omega)$, is related to the antisymmetrized transition matrix elements by the
usual relation ($\hbar =1$)
\begin{equation}
M ({\bm \kappa}^{\prime},{\bm \kappa},\omega) = \braket{{\bm \kappa}^{\prime}|M (\omega)|{\bm \kappa}}
= - 4 \pi^2 \mu \braket{{\bm \kappa}^{\prime}|t (\omega)|{\bm \kappa}} \, ,
\end{equation}
where $\mu$ is the $NN$ reduced mass. The most general form of this amplitude, consistent with invariance under rotation, time reversal,
and parity is \cite{PhysRev.85.947}
\begin{equation}\label{nnamplitude}
\begin{split}
M &= a + c ({\bm \sigma}_1 + {\bm \sigma}_2 ) \cdot \hat{{\bm n}}
+ m ({\bm \sigma}_1 \cdot \hat{{\bm n}}) ({\bm \sigma}_2 \cdot \hat{{\bm n}}) \\
&+ (g+h) ({\bm \sigma}_1 \cdot \hat{{\bm l}}) ({\bm \sigma}_2 \cdot \hat{{\bm l}})
+ (g-h) ({\bm \sigma}_1 \cdot \hat{{\bm m}}) ({\bm \sigma}_2 \cdot \hat{{\bm m}}) \, ,
\end{split}
\end{equation}
where
\begin{equation}
\hat{{\bm l}} = \frac{{\bm \kappa}^{\prime} + {\bm \kappa}}{|{\bm \kappa}^{\prime} + {\bm \kappa}|} \, , \qquad
\hat{{\bm m}} = \frac{{\bm \kappa}^{\prime} - {\bm \kappa}}{|{\bm \kappa}^{\prime} - {\bm \kappa}|} \, , \qquad
\hat{{\bm n}} = \frac{{\bm \kappa} \times {\bm \kappa}^{\prime}}{|{\bm \kappa} \times {\bm \kappa}^{\prime}|} \, ,
\end{equation}
are the unit vectors defined by the $NN$ scattering plane. The amplitudes $a$, $c$, $m$, $g$, and $h$ can be expressed as complex
functions of $\omega$, ${\bm \kappa}$, and ${\bm \kappa}^{\prime}$. The amplitudes in Eq.~(\ref{nnamplitude}) are given in the
Hoshizaki notation \cite{Hoshizaki}. There are different ways to define them, a survey of the other decompositions can be found in
Refs.~\cite{Bystricky,Moravcsik} and references therein. We may also note that for an even-even nucleus with $J=0$, terms linear in the
spin of the target nucleons average to zero; only $a$ and $c$ amplitudes survive and they are connected to the central and spin-orbit
part of the $NN$ $t$ matrix, respectively.

The $NN$ amplitudes are usually expressed in terms of the decomposition of the scattering amplitude into
components describing spin singlet $(S=0)$ and spin triplet $(S=1)$ scattering, $M_{\nu^{\prime} \nu}^{S}$, where $\nu$
and $\nu^{\prime}$ refer to the incident and final spin projections in the triplet state. In the representation in which
these projections are referred to an axis of quantization along the incident beam direction $({\bm \kappa})$ we have
\begin{equation}
a = \frac{1}{4} (2 M_{11}^1 + M_{00}^1 + M_{00}^0) \, ,
\end{equation}
\begin{equation}
c = \frac{i}{2 \sqrt{2}} (M_{10}^1 - M_{01}^1) \, ,
\end{equation}
\begin{equation}
m = \frac{1}{4} (M_{00}^1 - 2 M_{1 -1}^1 - M_{00}^0) \, ,
\end{equation}
\begin{equation}
g = \frac{1}{4} (M_{11}^1 - M_{00}^0 + M_{1 -1}^1) \, ,
\end{equation}
\begin{equation}
h = \frac{1}{4 \cos \phi} (M_{11}^1 - M_{00}^1 - M_{1 -1}^1) \, .
\end{equation}
The amplitudes $M_{\nu^{\prime} \nu}^{S} = \braket{{\bm \kappa} S \nu^{\prime}|M (\omega)|{\bm \kappa} S \nu}$ and hence
$a$-$h$, are obtained \cite{PhysRevC.29.2267,PhysRevC.46.279} in terms of the partial wave components of the $NN$
amplitude, $M_{L^{\prime}L}^{JS} (\kappa^{\prime},\kappa;\omega)$, defined by
\begin{equation}
\begin{split}
M ({\bm \kappa}^{\prime},{\bm \kappa};\omega) = \frac{2}{\pi} \sum_{JLL^{\prime}SM} &i^{L-L^{\prime}}
\mathcal{Y}_{JM}^{L^{\prime}S} ({\hat {\bm \kappa}}^{\prime}) \\
&\times M_{L^{\prime}L}^{JS} (\kappa^{\prime},\kappa;\omega) \,
\mathcal{Y}_{JM}^{LS\, \dagger} ({\hat {\bm \kappa}}) \, ,
\end{split}
\end{equation}
where $\mathcal{Y}_{JM}^{LS}$ is the spin-angular function
\begin{equation}\label{spinangularfun}
\mathcal{Y}_{JM}^{LS} ({\hat {\bm \kappa}}) = \sum_{\Lambda \nu} (L \Lambda S \nu | J M) \,
Y_L^{\Lambda} ({\hat {\bm \kappa}}) \otimes \chi_{S \nu} \, ,
\end{equation}
and $Y_L^{\Lambda}$ and $\chi_{S \nu}$ are the spherical harmonic and the spin wave function of the $NN$ pair, respectively.
Explicitly, we have
\begin{equation}
\begin{split}
M_{\nu^{\prime}\nu}^S &= \frac{2}{\pi} \sum_{JMLL^{\prime}\Lambda \Lambda^{\prime}} i^{L-L^{\prime}}
(L^{\prime} \Lambda^{\prime} S \nu^{\prime} |J M) (L \Lambda S \nu | J M) \\
&\times Y_{L^{\prime}}^{\Lambda^{\prime}} ({\hat {\bm \kappa}}^{\prime})
Y_L^{\Lambda \, \ast} ({\hat {\bm \kappa}}) \, M_{L^{\prime}L}^{JS} (\kappa^{\prime},\kappa;\omega) \, .
\end{split}
\end{equation}

Detailed formulas for the required $M_{\nu^{\prime} \nu}^{S}$ amplitudes in terms of the partial wave amplitudes
$M_{L^{\prime}L}^{JS} (\kappa^{\prime},\kappa;\omega)$ for a quantization axis along the incident beam direction can be found
in Refs.~\cite{PhysRevC.29.2267,PhysRevC.46.279}. According to these formulae, the $a$ and $c$ amplitudes are given by
\begin{equation}\label{amplitudea}
\begin{split}
a_{pN} &= \frac{1}{f_{pN} \pi^2} \sum_{L=0}^{\infty} P_L (\cos \phi) \, \Big[ (2 L + 1) \, M_{LL}^{L,S=0} \\
&+ (2 L + 1) \, M_{LL}^{L,S=1} + (2 L + 3) \, M_{LL}^{L+1,S=1} \\
&+ (2 L - 1) \, M_{LL}^{L-1,S=1} \Big] \, ,
\end{split}
\end{equation}
\begin{equation}\label{amplitudec}
\begin{split}
c_{pN} &= \frac{i}{f_{pN} \pi^2} \sum_{L=1}^{\infty} P_L^1 (\cos \phi) \, \bigg[ \bigg( \frac{2L+3}{L+1} \bigg) \, M_{LL}^{L+1,S=1} \\
&- \bigg( \frac{2L+1}{L(L+1)} \bigg) \, M_{LL}^{L,S=1} - \bigg( \frac{2L-1}{L} \bigg) \, M_{LL}^{L-1,S=1} \bigg] \, ,
\end{split}
\end{equation}
where $f_{pp}=4$, $f_{pn}=8$, and $P_L^1 (x)$ are the associated Legendre polynomials
\begin{equation}
P_L^1 (x) = \sqrt{1-x^2} \, \frac{d}{d x} P_L (x) \, .
\end{equation}

\begin{table}[t]
\begin{center}
{\renewcommand{\arraystretch}{2.0}
\begin{tabular}{|l|l|}
\hline
$t_{L,LL}^{S=0,T=1} \hspace{0.3cm} : \quad {}^1S_0 , {}^1D_2 , {}^1G_4 , {}^1I_6 , {}^1K_8$ \\
\hline
$t_{L-1,LL}^{S=1,T=1} \hspace{0.3cm} : \quad {}^3P_0 , {}^3F_2 , {}^3H_4 , {}^3J_6 , {}^3L_8$ \\
\hline
$t_{L,LL}^{S=1,T=1} \hspace{0.3cm} : \quad {}^3P_1 , {}^3F_3 , {}^3H_5 , {}^3J_7$ \\
\hline
$t_{L+1,LL}^{S=1,T=1} \hspace{0.3cm} : \quad {}^3P_2 , {}^3F_4 , {}^3H_6 , {}^3J_8$ \\
\hline
$t_{L,LL}^{S=0,T=0} \hspace{0.3cm} : \quad {}^1P_1 , {}^1F_3 , {}^1H_5 , {}^1J_7$ \\
\hline
$t_{L-1,LL}^{S=1,T=0} \hspace{0.3cm} : \quad {}^3D_1 , {}^3G_3 , {}^3I_5 , {}^3K_7$ \\
\hline
$t_{L,LL}^{S=1,T=0} \hspace{0.3cm} : \quad {}^3D_2 , {}^3G_4 , {}^3I_6 , {}^3K_8$ \\
\hline
$t_{L+1,LL}^{S=1,T=0} \hspace{0.3cm} : \quad {}^3S_1 , {}^3D_3 , {}^3G_5 , {}^3I_7$ \\
\hline
\end{tabular}
}
\caption{\label{pwtmat} Partial waves of the $NN$ potential used to construct the three-dimensional $NN$ $t$ matrix
$t_{pN} ({\bm \kappa}^{\prime},{\bm \kappa};\omega)$.}
\end{center}
\end{table}

From these equations we can obtain the explicit expressions for the {\it pp} and {\it pn} central and
spin-orbit parts of the $NN$ $t$ matrix. As stated above, the optical potential is an operator in the spin space of the projectile and
the spin dependence is made explicit writing the $t$ matrix in the form ($N=p,n$):
\begin{equation}
t_{pN} ({\bm \kappa}^{\prime},{\bm \kappa};\omega) = t_{pN}^c ({\bm \kappa}^{\prime},{\bm \kappa};\omega)
+ \frac{i}{2} {\bm \sigma} \cdot {\bm \kappa}^{\prime} \times {\bm \kappa} \;
t_{pN}^{ls} ({\bm \kappa}^{\prime},{\bm \kappa};\omega) \, .
\end{equation}
In terms of the partial wave components $t_{JLL}^{ST} (\kappa^{\prime},\kappa;\omega)$ we have the following results for the
central part:
\begin{equation}
\begin{split}
t_{pp}^c &= \frac{1}{4 \pi^2} \sum_{L=0}^{\infty} P_L (\cos \phi) \,
\Big[ (2 L + 1) \, t_{L,LL}^{S=0,T=1} \\
&+ (2 L + 1) \, t_{L,LL}^{S=1,T=1} 
+ (2 L - 1) \, t_{L-1,LL}^{S=1,T=1} \\
&+ (2 L + 3) \, t_{L+1,LL}^{S=1,T=1} \Big] \, ,
\end{split}
\end{equation}
\begin{equation}
\begin{split}
t_{pn}^c &= \frac{1}{8 \pi^2} \sum_{L=0}^{\infty} P_L (\cos \phi) \,
\Big[ (2 L + 1) \, t_{L,LL}^{S=0,T=0} \\
&+ (2 L + 1) \, t_{L,LL}^{S=1,T=0}
+ (2 L - 1) \, t_{L-1,LL}^{S=1,T=0} \\
&+ (2 L + 3) \, t_{L+1,LL}^{S=1,T=0}
+ (2 L + 1) \, t_{L,LL}^{S=0,T=1} \\
&+ (2 L + 1) \, t_{L,LL}^{S=1,T=1}
+ (2 L - 1) \, t_{L-1,LL}^{S=1,T=1} \\
&+ (2 L + 3) \, t_{L+1,LL}^{S=1,T=1} \Big] \, .
\end{split}
\end{equation}
and, similarly, for the spin-orbit part:
\begin{equation}
\begin{split}
t_{pp}^{ls} &= - \frac{1}{2 \pi^2} \sum_{L=1}^{\infty}
\frac{d \, P_L (\cos \phi)}{d \cos \phi} \, \frac{1}{\kappa^{\prime}\kappa}
\bigg[ - \frac{2 L - 1}{L} \, t_{L-1,LL}^{S=1,T=1} \\
&- \frac{2 L + 1}{L(L+1)} \, t_{L,LL}^{S=1,T=1}
+ \frac{2 L + 3}{L+1} \, t_{L+1,LL}^{S=1,T=1} \bigg] \, ,
\end{split}
\end{equation}
\begin{equation}
\begin{split}
t_{pn}^{ls} &= - \frac{1}{4 \pi^2} \sum_{L=1}^{\infty}
\frac{d \, P_L (\cos \phi)}{d \cos \phi} \, \frac{1}{\kappa^{\prime}\kappa}
\bigg[ - \frac{2 L - 1}{L} \, t_{L-1,LL}^{S=1,T=0} \\
&- \frac{2 L + 1}{L(L+1)} \, t_{L,LL}^{S=1,T=0}
+ \frac{2 L + 3}{L+1} \, t_{L+1,LL}^{S=1,T=0} \\
&- \frac{2 L - 1}{L} \, t_{L-1,LL}^{S=1,T=1}
- \frac{2 L + 1}{L(L+1)} \, t_{L,LL}^{S=1,T=1} \\
&+ \frac{2 L + 3}{L+1} \, t_{L+1,LL}^{S=1,T=1} \bigg] \, .
\end{split}
\end{equation}
The partial wave components $t_{JLL}^{ST} (\kappa^{\prime},\kappa;\omega)$ are computed in the $NN$ center-of-mass frame,
from the $NN$ potential. In this work we use two different
versions of the chiral potential at the fourth order (\n3lo) developed by Entem and Machleidt~\cite{chiralmachleidt} and
Epelbaum, Gl\"ockle, and Mei\ss ner~\cite{chiralepelbaum} and the CD-Bonn~\cite{cdbonn} potential.
The $t_{JLL}^{ST} (\kappa^{\prime},\kappa;\omega)$ matrices are
computed for each partial waves up to $J=8$. The partial waves are collected in the Tab.~\ref{pwtmat}.

\subsection{The Transition Amplitude in the Partial Wave Representation}
\label{transparwave}
The optimally factorized first-order KMT optical potential as an operator in the spin space of the projectile is given
in Eq.~(\ref{opticaldec}) as
\begin{equation}\label{decu}
\hat{U} ({\bm k}^{\prime},{\bm k};\omega) =  \hat{U}^c ({\bm k}^{\prime},{\bm k};\omega)
+ \frac{i}{2} {\bm \sigma} \cdot {\bm k}^{\prime} \times {\bm k} \; \hat{U}^{ls} ({\bm k}^{\prime},{\bm k};\omega) \, .
\end{equation}
From the conservation of the total angular momentum and parity, this spin operator can be expanded as
\begin{equation}\label{naexp}
\hat{U} ({\bm k}^{\prime},{\bm k};\omega) = \frac{2}{\pi} \sum_{JLM} \mathcal{Y}_{JM}^{L\frac{1}{2}} ({\hat {\bm k}}^{\prime}) \,
\hat{U}_{LJ} (k^{\prime},k;\omega) \, \mathcal{Y}_{JM}^{L\frac{1}{2}\, \dagger} ({\hat {\bm k}}) \, ,
\end{equation}
where $J=L \pm 1/2$ and $\mathcal{Y}_{JM}^{L\frac{1}{2}}$ is the standard spin-angular function of Eq.~(\ref{spinangularfun})
Inserting the expansion in Eq.~(\ref{naexp}) into the Eq.~(\ref{lipschw}) we obtain the same decomposition for
the $T$ matrix
\begin{equation}
\hat{T} ({\bm k}^{\prime},{\bm k};E) = \frac{2}{\pi} \sum_{JLM} \mathcal{Y}_{JM}^{L\frac{1}{2}} ({\hat {\bm k}}^{\prime}) \,
\hat{T}_{LJ} (k^{\prime},k;E) \, \mathcal{Y}_{JM}^{L\frac{1}{2}\, \dagger} ({\hat {\bm k}}) \, ,
\end{equation}
where the partial-wave components of the transition operator for the elastic scattering are given by
\begin{equation}\label{tlj}
\begin{split}
\hat{T}_{LJ} (k^{\prime},k;E) &= \hat{U}_{LJ} (k^{\prime},k;\omega) \\
&+ \frac{2}{\pi} \int_0^{\infty} d p \, p^2 \frac{\hat{U}_{LJ} (k^{\prime},p;\omega) \,
\hat{T}_{LJ} (p,k;E)}{E (k_0) - E (p) + i \epsilon} \, ,
\end{split}
\end{equation}
where
\begin{equation}
E (k_0) = \sqrt{k_0^2 + m_{proj}^2} + \sqrt{k_0^2 + m_{targ}^2} \, ,
\end{equation}
\begin{equation}
E (p) = \sqrt{p^2 + m_{proj}^2} + \sqrt{p^2 + m_{targ}^2} \, ,
\end{equation}
and $m_{proj}$ and $m_{targ}$ are the masses of the projectile and of the target, respectively. In terms of the partial wave components
of the quantities $\hat{U}^c ({\bm k}^{\prime},{\bm k};\omega)$ and $\hat{U}^{ls} ({\bm k}^{\prime},{\bm k};\omega)$, we have
\begin{equation}\label{pwu}
\hat{U}_{LJ} (k^{\prime},k;\omega) = \hat{U}_L^c (k^{\prime},k;\omega) + C_{LJ} \, \hat{V}_L^{ls} (k^{\prime},k;\omega) \, , 
\end{equation}
where
\begin{equation}
\begin{split}
C_{LJ} &= \frac{1}{2} \left[ J(J+1) - L(L+1) - \frac{3}{4} \right] \, , \\
\hat{V}_L^{ls} (k^{\prime},k;\omega) &= \frac{k^{\prime}k}{2L+1} \left[ \hat{U}_{L+1}^{ls} (k^{\prime},k;\omega) - \hat{U}_{L-1}^{ls} (k^{\prime},k;\omega) \right] \, .
\end{split}
\end{equation}
To obtain these results, the quantities $\hat{U}^c ({\bm k}^{\prime},{\bm k};\omega)$ and $\hat{U}^{ls} ({\bm k}^{\prime},{\bm k};\omega)$ are
expanded in a manner similar to Eq.~(\ref{naexp}), with the difference that the partial wave components, $\hat{U}_L^c$ and
$\hat{U}_L^{ls}$, are independent of $J$.

The partial wave components of $\hat{U}^c ({\bm k}^{\prime},{\bm k};\omega)$ and $\hat{U}^{ls} ({\bm k}^{\prime},{\bm k};\omega)$ can be
calculated in terms of the $NN$ $t$-matrix components and of the nuclear densities from Eq.~(\ref{opticalcentral}).
The projection can be performed numerically by evaluating the integral
\begin{equation}
\begin{split}
\hat{U}_L^a (k^{\prime},k;\omega) &= \pi^2 \int_{-1}^{+1} d x \, P_L (x) \hat{U}^a ({\bm k}^{\prime},{\bm k};\omega) \\
&= \pi^2 \int_{-1}^{+1} d x \, P_L (x) \hat{U}^a (k^{\prime},k,x;\omega) \, ,
\end{split}
\end{equation}
where $x=\cos \theta$ and the potentials in terms of $k^{\prime}$ and $k$ are obtained from Eq.~(\ref{opticalcentral}) with
\begin{equation}
\begin{split}
q (x) &= \sqrt{k^{\prime \, 2} + k^2 - 2 k^{\prime} k x} \, , \\
K (x) &= \frac{1}{2} \sqrt{k^{\prime \, 2} + k^2 + 2 k^{\prime} k x} \, , \\
{\bm q} \cdot {\bm K} &= \frac{1}{2} \left( k^{\prime \, 2} - k^2 \right) \, .
\end{split}
\end{equation}
The one-dimensional integral equation for the partial wave elements $\hat{T}_{LJ}$, Eq.~(\ref{tlj}), is solved for the complex
potentials $\hat{U}_{LJ}$. In actual calculations, the number of $L$ values needed to represent the nuclear optical potential at
the level of accuracy required through the partial wave components $\hat{U}_{LJ} ({\bm k}^{\prime},{\bm k};\omega)$ can be as large
as $30$ for a $^{16}$O target at $200$ MeV.

\subsection{The Scattering Observables}
\label{scatobs}
Under the assumptions of parity conservation and rotational invariance, the most general form of the full amplitude for the elastic
proton scattering from a spin $0$ nucleus is given by
\begin{equation}\label{ampm}
M (k_0 ,\theta) = A (k_0 ,\theta) + {\bm \sigma} \cdot \hat{{\bm N}} \, C (k_0 ,\theta) \, ,
\end{equation}
where the amplitudes $A (k_0 ,\theta)$ and $C (k_0 ,\theta)$ are obtained from the partial wave solutions of Eq.~(\ref{tlj}) as
\begin{equation}\label{naamplitudea}
A (\theta) = \frac{1}{2 \pi^2} \sum_{L=0}^{\infty} \left[ (L+1) F_L^+ (k_0) + L F_L^- (k_0) \right] P_L (\cos \theta) \, , \\
\end{equation}
\begin{equation}\label{naamplitudec}
C (\theta) = \frac{i}{2 \pi^2} \sum_{L=1}^{\infty} \left[ F_L^+ (k_0) - F_L^- (k_0) \right] P_L^1 (\cos \theta) \, .
\end{equation}
In Eqs.~(\ref{naamplitudea}) and (\ref{naamplitudec}) an implicit dependence on $k_0$ is assumed.
The functions $F_L^{\pm}$ denote $F_{LJ}$ for $J=L\pm 1/2$, respectively, and are given as
\begin{equation}
F_{LJ} (k_0) = - \frac{A}{A-1} 4 \pi^2 \mu (k_0) \hat{T}_{LJ} (k_0 , k_0 ; E) \, ,
\end{equation}
where the relativistic reduced mass is
\begin{equation}\label{nareducedmass}
\mu (k_0) = \frac{E_{proj} (k_0) \, E_{targ} (k_0)}{E_{proj} (k_0) + E_{targ} (k_0)} \, .
\end{equation}
Three independent scattering observable can be considered: the unpolarized differential cross section,
the analyzing power $A_y$, and the spin rotation $Q$. Their expressions as functions of the the amplitudes $A$ and $C$ are:
\begin{equation}\label{diffcrossec}
\frac{d \sigma}{d \Omega} (\theta) = {|A (\theta)|}^2 + {| C (\theta) |}^2 \, ,
\end{equation}
\begin{equation}\label{analyzingpower}
A_y (\theta) = \frac{2 \mathrm{Re} [A^{\ast} (\theta) \, C (\theta)]}{{|A (\theta)|}^2 + {| C (\theta) |}^2} \, ,
\end{equation}
\begin{equation}\label{spinrotation}
Q (\theta) = \frac{2 \mathrm{Im} [A (\theta) \, C^{\ast} (\theta)]}{{|A (\theta)|}^2 + {|C (\theta)|}^2} \, .
\end{equation}

\subsection{Treatment of the Coulomb Potential}
\label{coulpot}
In this section we include in the theoretical framework the Coulomb interaction between the incoming proton, with charge $e$, and the
spin $0$ target, with charge $Z e$. This has been done following the algorithm outlined in Refs.~\cite{PhysRevC.44.1569,elster1993}.
The interaction is separated into the sum of two parts: the ``point'' Coulomb interaction and the short-ranged one which is given
by the sum of the nuclear potential and the short-range Coulomb interaction due to the finite dimension of the nucleus.
Since the Coulomb $T$-matrix is known analytically, we only need to compute the transition matrix modified by the residual Coulomb
field.

In this approach the total scattering amplitude can be written in the standard way as
\begin{equation}
M (k_0 ,\theta) = A (k_0 ,\theta) + {\bm \sigma} \cdot \hat{{\bm N}} \, C (k_0 ,\theta) \, ,
\end{equation}
where now instead of Eqs.~(\ref{naamplitudea}) and (\ref{naamplitudec}) we have
\begin{equation}\label{coulamplitudea}
\begin{split}
A (k_0 ,\theta) &= F_{pt}^c (k_0 ,\theta) + \frac{1}{2 \pi^2} \sum_{L=0}^{\infty} e^{2 i \sigma_L} \big[ (L+1) \bar{F}_L^+ (k_0) \\
&+ L \bar{F}_L^- (k_0) \big] P_L (\cos \theta) \, , \\
\end{split}
\end{equation}
\begin{equation}\label{coulamplitudec}
C (k_0 ,\theta) = \frac{i}{2 \pi^2} \sum_{L=1}^{\infty} e^{2 i \sigma_L} \big[ \bar{F}_L^+ (k_0) - \bar{F}_L^- (k_0) \big] P_L^1 (\cos \theta) \, .
\end{equation}
In Eqs.~(\ref{coulamplitudea}) and (\ref{coulamplitudec}) $F_{pt}^c (k_0 ,\theta)$ is the Coulomb scattering amplitude due to a point
charge \cite{rodberg1967}
\begin{equation}
F_{pt}^c (k_0 ,\theta) = \frac{- \eta (k_0) \, \exp \big[ 2 i \sigma_0 - i \eta (k_0) \ln (1-\cos \theta) \big]}{k_0 (1-\cos \theta)} \, ,
\end{equation}
where
\begin{equation}\label{sommerfeld}
\eta (k) = \frac{\mu Z \alpha}{k}
\end{equation}
is the Sommerfeld parameter, $\mu$ is the reduced mass of Eq.~(\ref{nareducedmass}), and $\alpha$ is the fine structure constant.
The Coulomb phase shifts $\sigma_L$ are given by
\begin{equation}
\sigma_L = \arg \Gamma \big[ L + 1 + i \eta (k_0) \big] \, .
\end{equation}

The partial wave scattering amplitudes $\bar{F}_L^{\pm}$ are obtained from the solution of the Coulomb distorted $\bar{T}$ matrix
\begin{equation}\label{tmatrixcoul}
\bar{T} ({\bm k}^{\prime},{\bm k};E) = \bar{U} ({\bm k}^{\prime},{\bm k};\omega)
+ \int d^3 p \frac{\bar{U} ({\bm k}^{\prime},{\bm p};\omega) \,
\bar{T} ({\bm p},{\bm k};E)}{E (k_0) - E (p) + i \epsilon} \, ,
\end{equation}
where
\begin{equation}\label{matelemencoul}
\begin{split}
\bar{U} ({\bm k}^{\prime},{\bm k};\omega) &= \braket{{\bm k}^{\prime}|\bar{U} (\omega)|{\bm k}}
= \braket{\psi_c^{(+)} ({\bm k}^{\prime})|\hat{U} (\omega)|\psi_c^{(+)} ({\bm k})} \, ,
\end{split}
\end{equation}
and $\psi_c^{(+)} ({\bm k})$ is the Coulomb distorted wave function.

In order to solve Eq.~(\ref{tmatrixcoul}), we need to be able to
generate the momentum space matrix element $\bar{U} ({\bm k}^{\prime},{\bm k};\omega)$ as given in Eq.~(\ref{matelemencoul}).
We begin with the potential $\hat{U} ({\bm k}^{\prime},{\bm k};\omega)$, discussed in Section~\ref{firstop}, and we transform it
into the coordinate space through the double Fourier transform
\begin{equation}\label{doublfourcoord}
\hat{U} ({\bm r}^{\prime},{\bm r};\omega) = \int d^3 k^{\prime} d^3 k \, \braket{{\bm r}^{\prime}|{\bm k}^{\prime}}
\hat{U} ({\bm k}^{\prime},{\bm k};\omega) \braket{{\bm k}|{\bm r}}
\end{equation}
and then we construct the matrix element of Eq.~(\ref{matelemencoul}) by folding $\hat{U} ({\bm r}^{\prime},{\bm r};\omega)$ with
coordinate space Coulomb wave functions
\begin{equation}\label{doublfourmomen}
\begin{split}
\bar{U} ({\bm k}^{\prime},{\bm k};\omega) = \int d^3 r^{\prime} d^3 r \, &\braket{\psi_c^{(+)} ({\bm k}^{\prime}) | {\bm r}^{\prime}} \\
&\times \hat{U} ({\bm r}^{\prime},{\bm r};\omega) \braket{{\bm r}|\psi_c^{(+)} ({\bm k})} \, .
\end{split}
\end{equation}
In the partial wave representation, Eq.~(\ref{doublfourcoord}) for the central and spin-orbit parts becomes
\begin{equation}\label{doublfourparwavecoor}
\begin{split}
\hat{U}_L^a (r^{\prime},r;\omega) &= \frac{4}{\pi^2} \int_0^{\infty} d k^{\prime} \, k^{\prime \, 2} \\
&\times \int_0^{\infty} d k \, k^2 j_L (k^{\prime} r^{\prime}) \hat{U}_L^a (k^{\prime},k;\omega) j_L (k r) \, ,
\end{split}
\end{equation}
where $j_L (k r)$ are the spherical Bessel functions. Similarly, Eq.~(\ref{doublfourmomen}) becomes
\begin{equation}\label{doublfourparwave}
\begin{split}
\bar{U}_L^a (k^{\prime},k;\omega) &= \frac{1}{k^{\prime} k} \int_0^{\infty} d r^{\prime} \, r^{\prime} \\
&\times \int_0^{\infty} d r \, r F_L (\eta , k^{\prime} r^{\prime}) \hat{U}_L^a (r^{\prime},r;\omega) F_L (\eta , k r) \, ,
\end{split}
\end{equation}
where $F_L$ is the regular Coulomb function.
The potential $\bar{U} ({\bm k}^{\prime},{\bm k};\omega)$ can be expanded in partial waves as in Eq.~(\ref{naexp})
\begin{equation}
\bar{U} ({\bm k}^{\prime},{\bm k};\omega) = \frac{2}{\pi} \sum_{JLM} \mathcal{Y}_{JM}^{L\frac{1}{2}} ({\hat {\bm k}}^{\prime}) \,
\bar{U}_{LJ} (k^{\prime},k;\omega) \, \mathcal{Y}_{JM}^{L\frac{1}{2}\, \dagger} ({\hat {\bm k}}) \, ,
\end{equation}
where
\begin{equation}\label{ubarlj}
\bar{U}_{LJ} (k^{\prime},k;\omega) = \bar{U}_L^c (k^{\prime},k;\omega) + C_{LJ} \, \bar{V}_L^{ls} (k^{\prime},k;\omega) \, , \\
\end{equation}
and 
\begin{equation}
\begin{split}
C_{LJ} &= \frac{1}{2} \left[ J(J+1) - L(L+1) - \frac{3}{4} \right] \, , \\
\bar{V}_L^{ls} (k^{\prime},k;\omega) &= \frac{k^{\prime}k}{2L+1} \left[ \bar{U}_{L+1}^{ls} (k^{\prime},k;\omega) - \bar{U}_{L-1}^{ls} (k^{\prime},k;\omega) \right] \, ,
\end{split}
\end{equation}
Likewise, we can expand the $\bar{T}$ matrix in Eq.~(\ref{tmatrixcoul}) as
\begin{equation}
\bar{T} ({\bm k}^{\prime},{\bm k};E) = \frac{2}{\pi} \sum_{JLM} \mathcal{Y}_{JM}^{L\frac{1}{2}} ({\hat {\bm k}}^{\prime}) \,
\bar{T}_{LJ} (k^{\prime},k;E) \, \mathcal{Y}_{JM}^{L\frac{1}{2}\, \dagger} ({\hat {\bm k}}) \, ,
\end{equation}
where the partial wave components are
\begin{equation}\label{tljbar}
\begin{split}
\bar{T}_{LJ} (k^{\prime},k;E) &= \bar{U}_{LJ} (k^{\prime},k;\omega) \\
&+ \frac{2}{\pi} \int_0^{\infty} d p \, p^2 \frac{\bar{U}_{LJ} (k^{\prime},p;\omega) \,
\bar{T}_{LJ} (p,k;E)}{E (k_0) - E (p) + i \epsilon} \, .
\end{split}
\end{equation}
The partial wave scattering amplitudes $\bar{F}_L^{\pm}$ entering Eqs.~(\ref{coulamplitudea}) and (\ref{coulamplitudec}) are
given by
\begin{equation}\label{fbaramp}
\bar{F}_{LJ} (k_0) = - \frac{A}{A-1} 4 \pi^2 \mu (k_0) \bar{T}_{LJ} (k_0 , k_0 ; E) \, .
\end{equation}

\section{The {\it NN} Amplitudes}
\label{nnresults}
\begin{figure}[t]
\begin{center}
\includegraphics[scale=0.33]{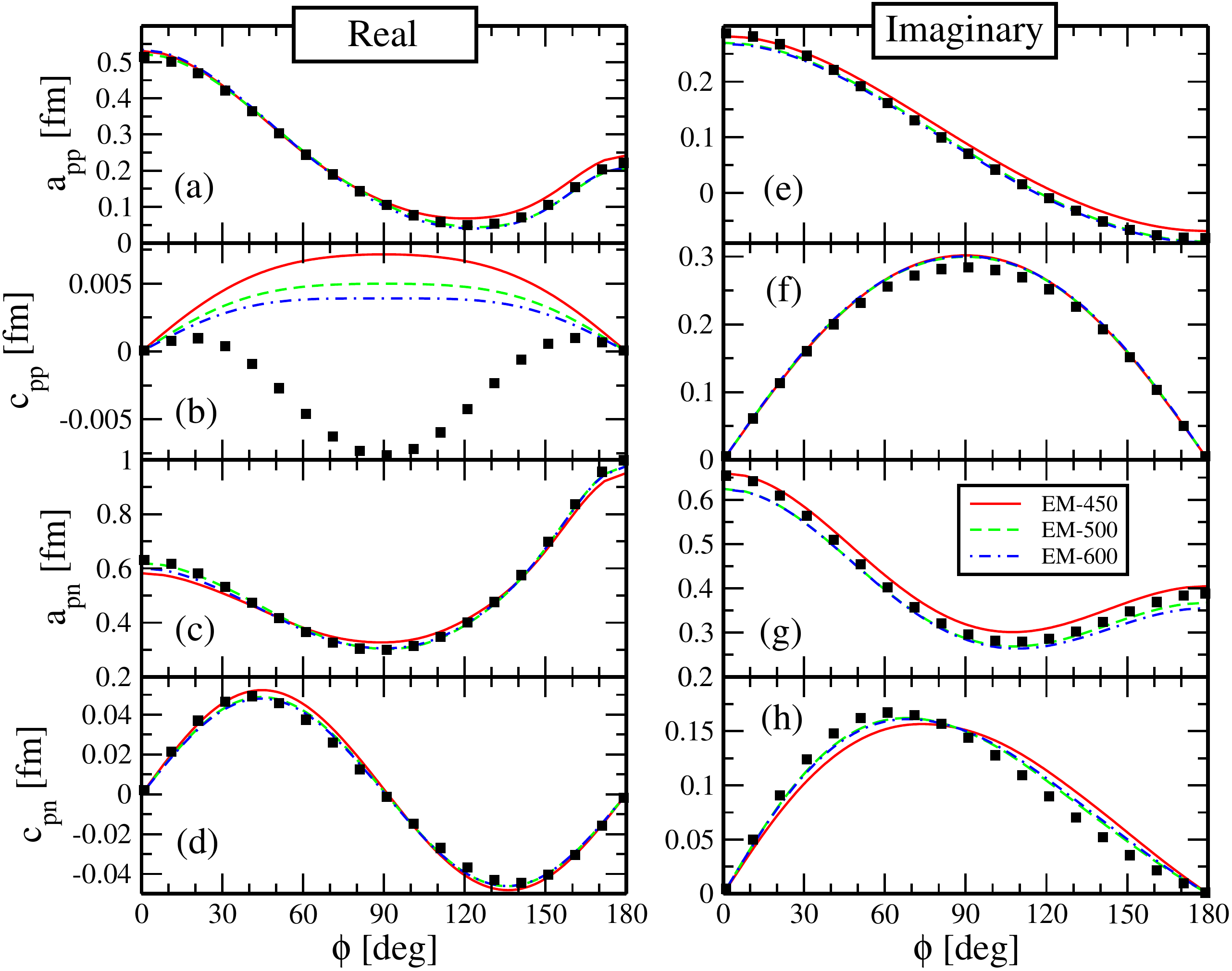}
\caption{\label{amp_100MeV_1}  (Color online) Real (left panel) and Imaginary (right panel) parts of {\it pp} and {\it pn} 
Wolfenstein amplitudes ($a$ and $c$) as functions of the center-of-mass $NN$ angle $\phi$. 
All the amplitudes are computed at $100$ MeV using the EM 
potentials \cite{chiralmachleidt,PhysRevC.88.054002,PhysRevC.87.014322,PhysRevC.75.024311} with a LS 
cutoff ranging between 450 and 600 MeV. Data (black squares) are taken from Ref.~\cite{nnonline}.}
\end{center}
\end{figure}
\begin{figure}[t]
\begin{center}
\includegraphics[scale=0.33]{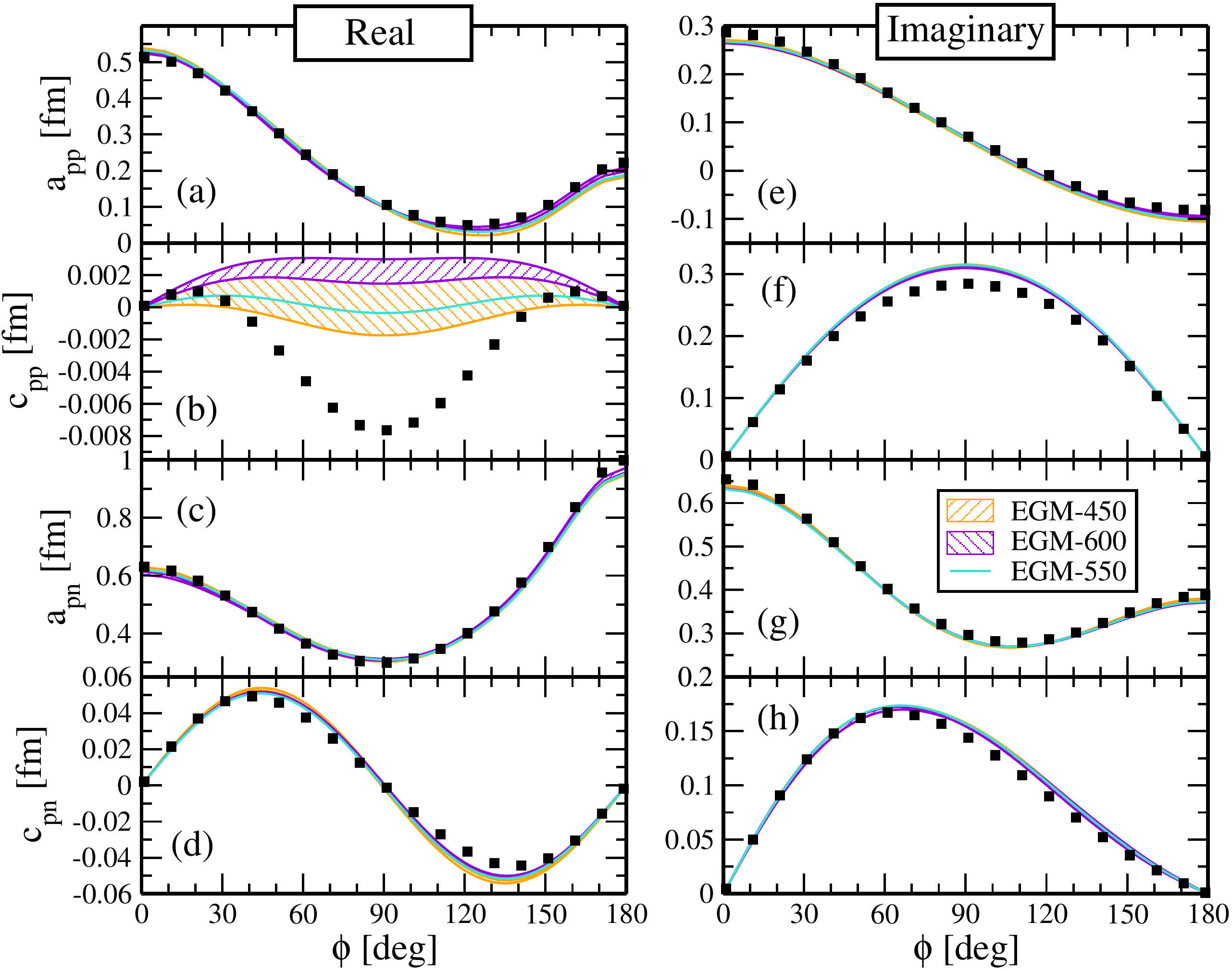}
\caption{\label{amp_100MeV_2}   (Color online) The same as in Fig.~\ref{amp_100MeV_1} 
using EGM potentials \cite{chiralepelbaum} with a LS 
cutoff ranging between 450 and 600 MeV. In two cases ($\Lambda = 450$ and $600$ MeV) we show uncertainty bands produced
by changing $\tilde{\Lambda}$ according to Eq. (\ref{cutoffs}).
Data (black squares) are taken from Ref.~\cite{nnonline}.}
\end{center}
\end{figure}
\begin{figure}[t]
\begin{center}
\includegraphics[scale=0.33]{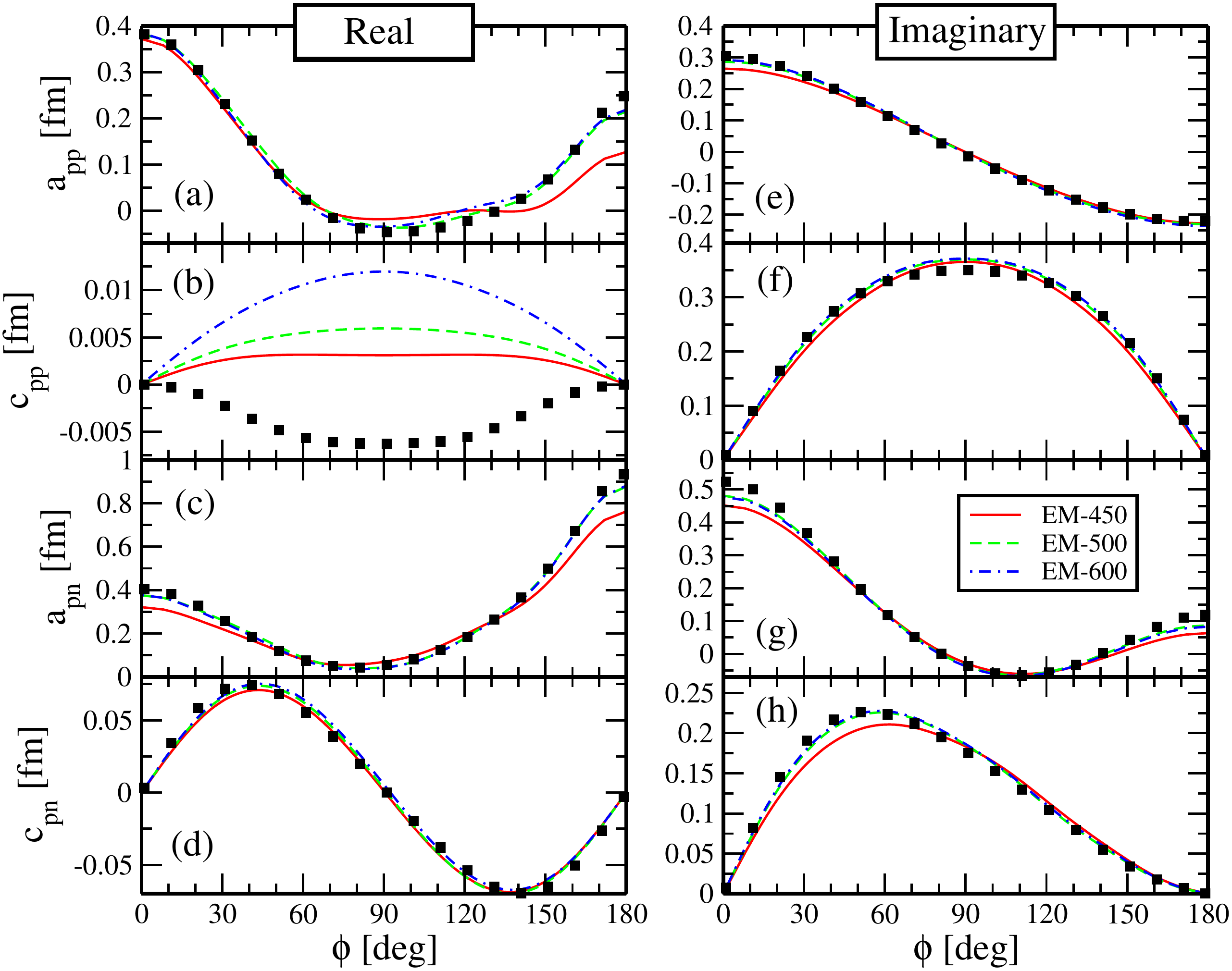}
\caption{\label{amp_200MeV_1}   (Color online) The same as in Fig.~\ref{amp_100MeV_1} but for an energy of $200$ MeV. Data (black squares) 
are taken from Ref.~\cite{nnonline}.}
\end{center}
\end{figure}
\begin{figure}[t]
\begin{center}
\includegraphics[scale=0.33]{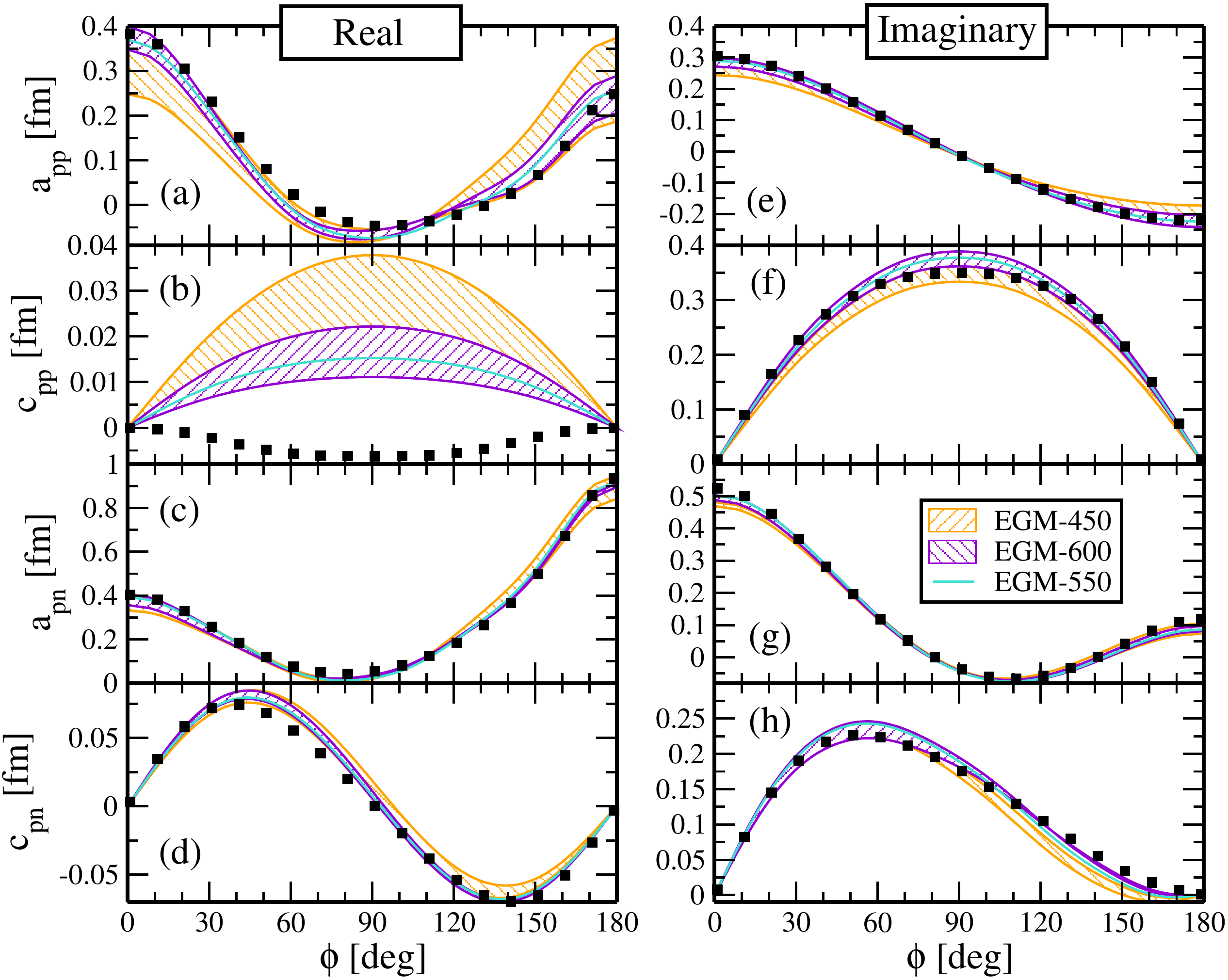}
\caption{\label{amp_200MeV_2} (Color online)  The same as is Fig.~\ref{amp_100MeV_2} but for an energy of $200$ MeV. Data (black squares)
are taken from Ref.~\cite{nnonline}.}
\end{center}
\end{figure}
\begin{figure}[t]
\begin{center}
\includegraphics[scale=0.33]{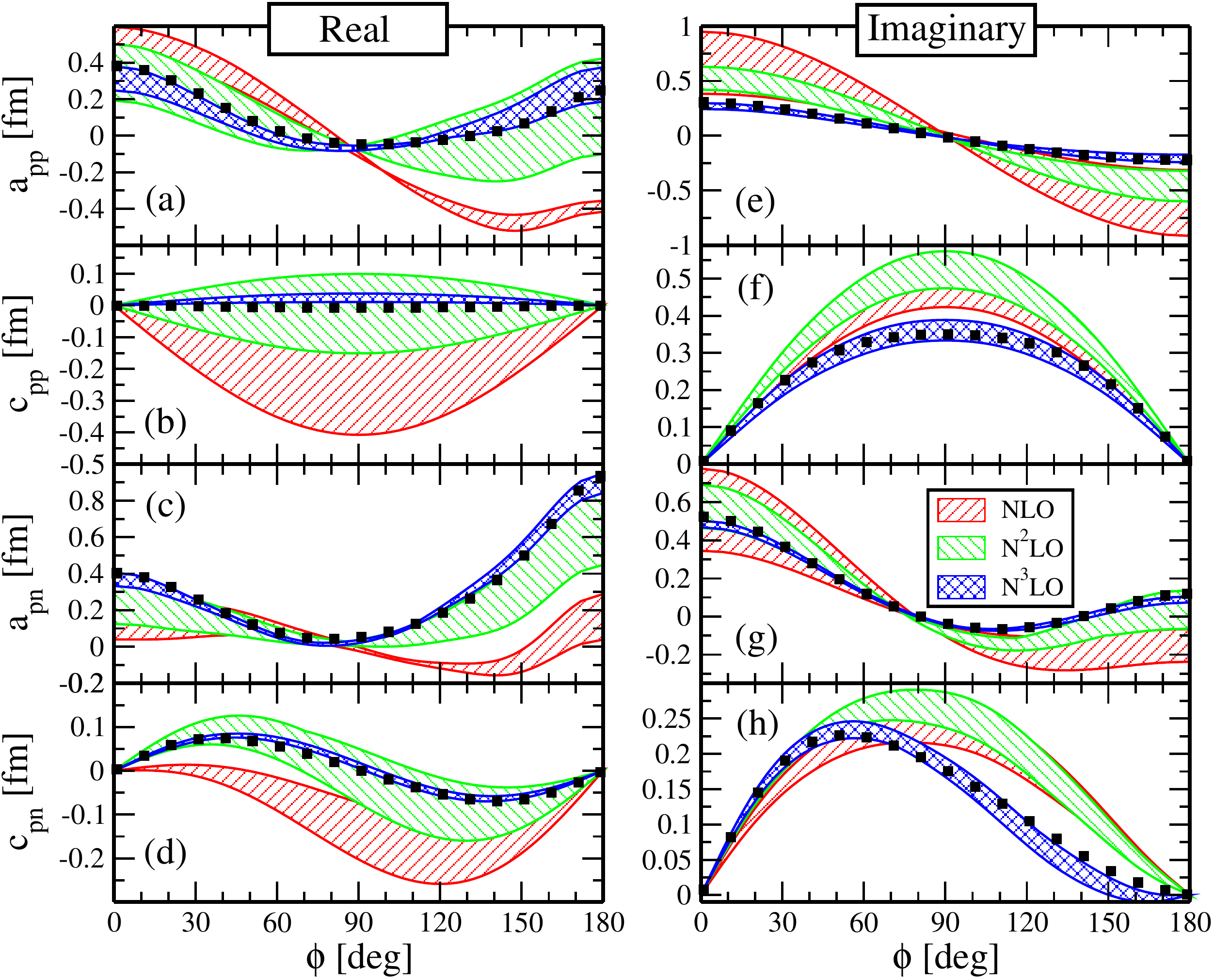}
\caption{\label{amp_200MeV_orders}   
(Color online) Real (left panel) and Imaginary (right panel) parts of {\it pp} and {\it pn} 
Wolfenstein amplitudes ($a$ and $c$) as functions of the center-of-mass $NN$ angle $\phi$ computed at $200$ MeV 
using the EGM potential \cite{chiralepelbaum} at different orders: 
red bands are the NLO results, green and blue bands are, respectively, the \3bn2lo and \n3lo
results.
Data (black squares) are taken from Ref.~\cite{nnonline}.}
\end{center}
\end{figure}

In this section we present and discuss the theoretical results for the {\it pp} and {\it pn} Wolfenstein amplitudes which are used to
compute the central $a$ (\ref{amplitudea}) and the spin-orbit part $c$ (\ref{amplitudec}) of the three-dimensional $NN$ $t$ matrix. 
Calculations are performed using two different versions of the chiral potential at fourth order (\n3lo)
based on the works of Entem and Machleidt~\cite{chiralmachleidt,PhysRevC.88.054002,PhysRevC.87.014322,PhysRevC.75.024311} and Epelbaum et {\it al.}~\cite{chiralepelbaum}. The performance of our code has been tested against the
CD-Bonn potential \cite{cdbonn} reproducing well known results  \cite{PhysRevC.57.1378,weppner} in order to check 
its numerical correctness.

Entem and Machleidt (EM), who first presented a chiral potential at the fourth order, treat divergent terms in the two-pion exchange (2PE)
contributions with dimensional regularization (DR), while Epelbaum, Gl\"ockle, and
Mei\ss ner (EGM) employ a spectral function regularization (SFR). 
In both cases the goal is to cut out the short-range part of the 2PE contribution that, as shown in Ref. \cite{PhysRevC.88.054002}, has unphysically strong attraction, particularly at \3bn2lo (for a comprehensive discussion about different regularization schemes we refer the reader to Sect. 3.2.1 of Ref. \cite{chiralepelbaum}).
As a usual procedure, the nucleon-nucleon potential entering the LS equation is multiplied by a regulator function $f^\Lambda$
\begin{equation}
V ({\bm k}, {\bm k}^{\prime}) \rightarrow V ({\bm k}, {\bm k}^{\prime}) f^\Lambda (k,k^{\prime})
\end{equation}
where
\begin{equation}
f^\Lambda =\exp \left( -(k^{\prime}/\Lambda)^{2n}  -(k/\Lambda)^{2n} \right) \quad {\rm with} \quad n=2,3\; .
\end{equation}

While Entem and Machleidt present results 
for three choices of the cutoff necessary to regulate the high-momentum components in the LS equation ($\Lambda=$ 450, 500, and 600 MeV), Epelbaum {\it et al.} ~\cite{chiralepelbaum}  allow also to study variations of the cutoff $\tilde{\Lambda}$ that regulates the 2PE contribution.
In fact, in the latter approach one can choose between the following cutoff combinations:
\begin{eqnarray}
\label{cutoffs}
\{\Lambda, \tilde{\Lambda} \} & = & \{450,500 \},  \{450,700 \},  \{550,600 \},  \nonumber\\
 ~ & ~ & \{600,600 \},  \{600,700 \} \; .
\end{eqnarray}

In the following figures all the results are labelled by an acronym (to distinguish the authors) followed by the value of the 
LS cutoff ($\Lambda$). In the EGM case, 
for $\Lambda = 450$ and $600$ MeV we plot bands that show how calculations
can change respect to variations of the SFR cutoff $\tilde{\Lambda}$.


In Fig.~\ref{amp_100MeV_1} the theoretical results for the real and imaginary parts of {\it pp} and {\it pn} 
Wolfenstein amplitudes ($a$ and $c$) computed at an energy of $100$ MeV are shown as functions of the 
center-of-mass $NN$ angle $\phi$ and compared with the experimental data.
The calculations are performed  using the EM 
potentials \cite{chiralmachleidt,PhysRevC.88.054002,PhysRevC.87.014322,PhysRevC.75.024311} with a LS 
cutoff ranging between 450 and 600 MeV.
The experimental data are globally 
reproduced by the three potentials, with the only remarkable exception of the real part of the $c_{pp}$ amplitude that is overestimated.
It must be considered, however, that this is a small quantity, i.e. two orders of magnitude smaller than the respective imaginary part, 
and it will only provide a very small contribution to the optical potential. 
Concerning the other amplitudes, some deviations from the experimental data are found, in particular for the imaginary part of the $c_{pp}$ and $c_{pn}$ amplitudes.
Finding some discrepancies is not surprising, because the
NN amplitudes are directly related to the empirical NN phase shifts, that are not always
perfectly reproduced by realistic potentials for some $J$ (see $^3F_3$, $^3F_4$, and $^3G_5$ cases in Figs. 8 and 9 of Ref. \cite{chiralmachleidt} and Fig. 27 of Ref. \cite{chiralepelbaum}).


In Fig.~\ref{amp_100MeV_2} we show the results obtained at $100$ MeV with the EGM potentials ~\cite{chiralepelbaum}.
Also in this case, all three potentials are in overall good agreement with the experimental data with the only remarkable
exception of the real part of the $c_{pp}$ amplitude. In particular, they show very similar results and in many cases the yellow and
torquoise bands are overlapped. Their trends are also very close to the ones shown in Fig.~\ref{amp_100MeV_1} for the EM potential
and they display the same discrepancy in comparison with the experimental data for $c_{pn}$ and around
the peak of the imaginary part of $c_{pp}$.


Since ChPT is a low-momentum expansion of QCD, we expect that, as the energy is increased, larger discrepancies appear respect to empirical data. In Figs.~\ref{amp_200MeV_1} and \ref{amp_200MeV_2} we present the results corresponding to the ones shown in Figs.~\ref{amp_100MeV_1} 
and \ref{amp_100MeV_2} but at an energy of $200$ MeV. 
As energy is increased, all potentials are still unable to reproduce the experimental data of the real part of the $c_{pp}$ amplitude, but
most of the chiral potentials give satisfactory results, in agreement with the data for all the other amplitudes, with one notable exception. 
In fact, in both approaches, potentials with a cutoff of 450 MeV (see Figs. \ref{amp_200MeV_1} and \ref{amp_200MeV_2})  fail 
to reproduce the real part of $a_{pp}$ and $a_{pn}$ and underestimate the imaginary part of the $c_{pn}$ amplitude. 
Based on these flaws, we predict an unsatisfactory result for a nucleon-nucleus optical potential if EM-450 or EGM-450 are employed at energies well above 100 MeV. 


Chiral potentials at orders NLO, \3bn2lo, and \n3lo have been constructed and compared in \cite{chiralepelbaum} for EGM and in \cite{PhysRevC.91.054311} for EM. As an example, in Fig.~\ref{amp_200MeV_orders} we present a systematic study order by order 
of the convergence pattern using the EGM potential. At each order, calculations have been performed for all the cutoff combinations in Eq.~(\ref{cutoffs}), the variation produced by the different combinations in the calculated amplitudes is depicted by the bands in the figure. From the results shown for all NN amplitudes, we can draw the conclusion that it is mandatory
to use potentials at order \n3lo. At orders  NLO and \3bn2lo the amplitudes not only underestimate or overestimate empirical data but also miss the overall shapes. 
The order by order convergence will be further explored in a forthcoming paper
using the recent \chiral4lo potential \cite{PhysRevLett.115.122301}.


\section{The Scattering Results}
\label{scattresults}

\begin{figure}[t]
\begin{center}
\includegraphics[scale=0.33]{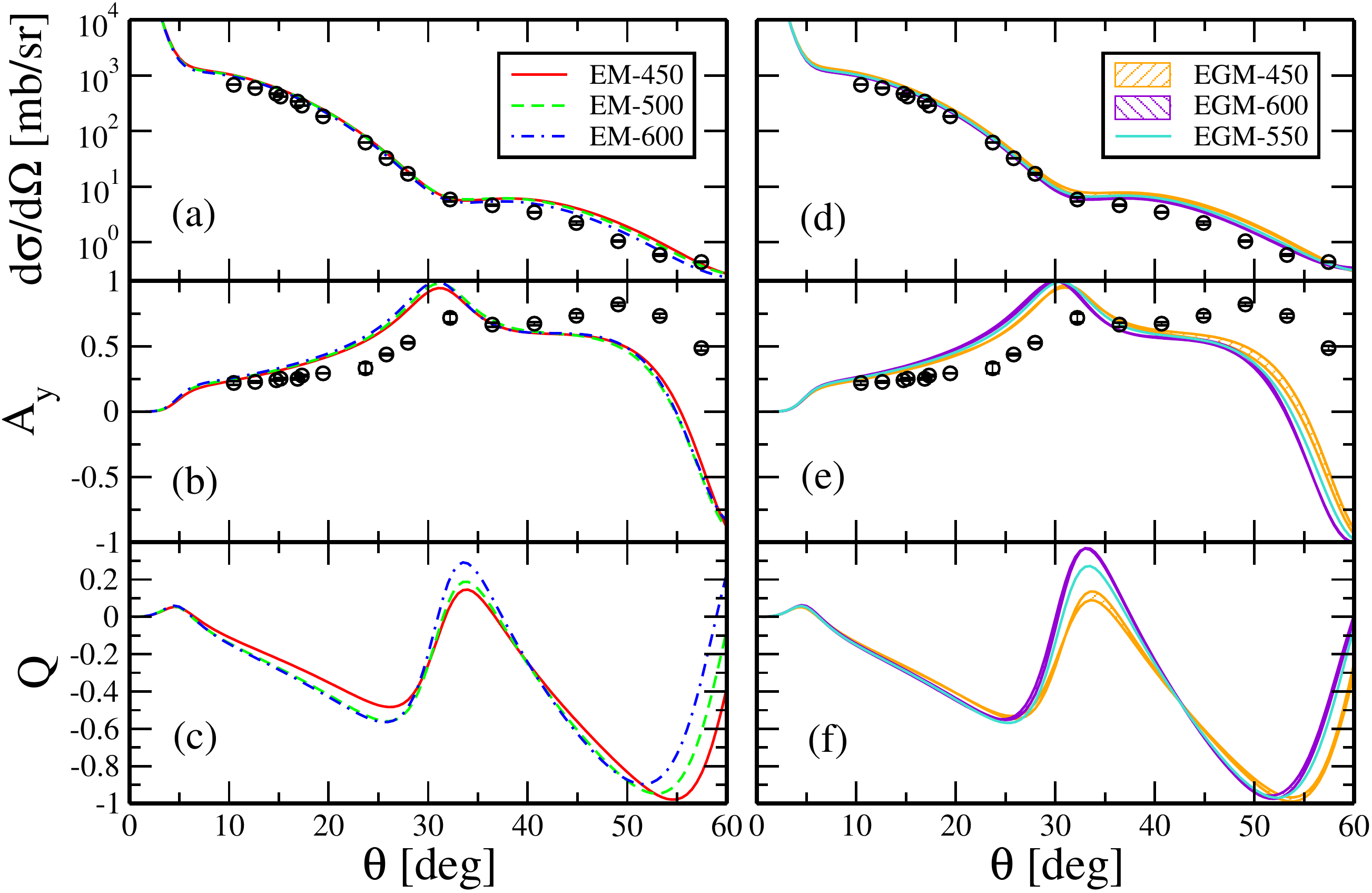}
\caption{\label{16O_100}  (Color online) Scattering observables (differential cross section $d\sigma/d\Omega$, 
analyzing power $A_y$, and spin rotation $Q$) as a function of the center-of-mass scattering angle $\theta$  
for elastic proton scattering on ${}^{16}$O computed at $100$ MeV
(laboratory energy). On the left panel we employ the set of EM 
potentials \cite{chiralmachleidt,PhysRevC.88.054002,PhysRevC.87.014322,PhysRevC.75.024311} 
while in the right panel we show the EGM potentials \cite{chiralepelbaum}. 
All potentials are denoted by the value of the LS cutoff.
Coulomb distortion is included as explained in Sect. \ref{coulpot}.
Data are taken from Refs.~\cite{kelly,exfor}.}
\end{center}
\end{figure}

\begin{figure}[t]
\begin{center}
\includegraphics[scale=0.33]{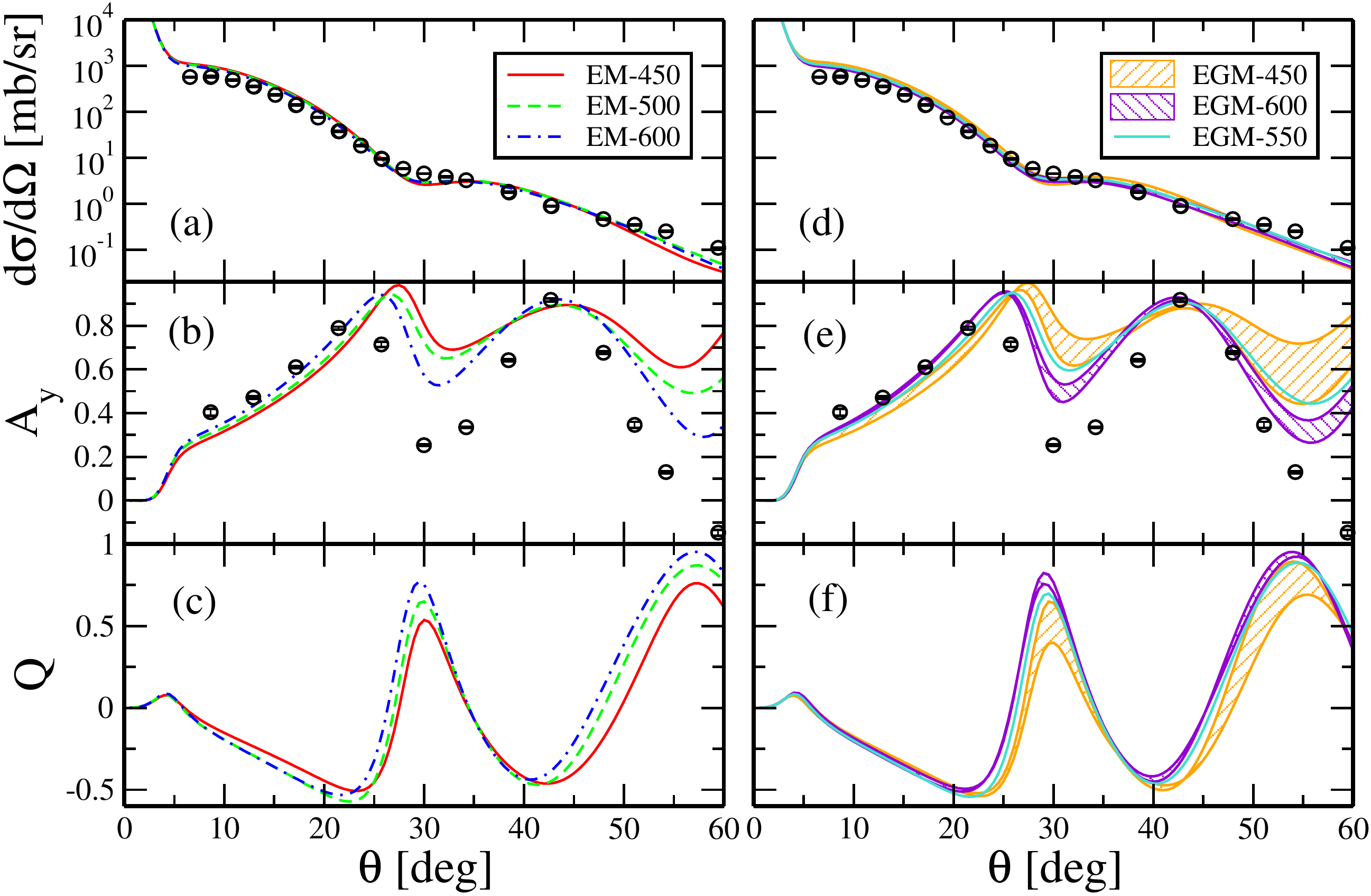}
\caption{\label{16O_135}  (Color online) The same as is Fig.~\ref{16O_100} but for an energy of $135$ MeV.
Data are taken from Refs.~\cite{kelly,exfor}.}
\end{center}
\end{figure}

\begin{figure}[t]
\begin{center}
\includegraphics[scale=0.33]{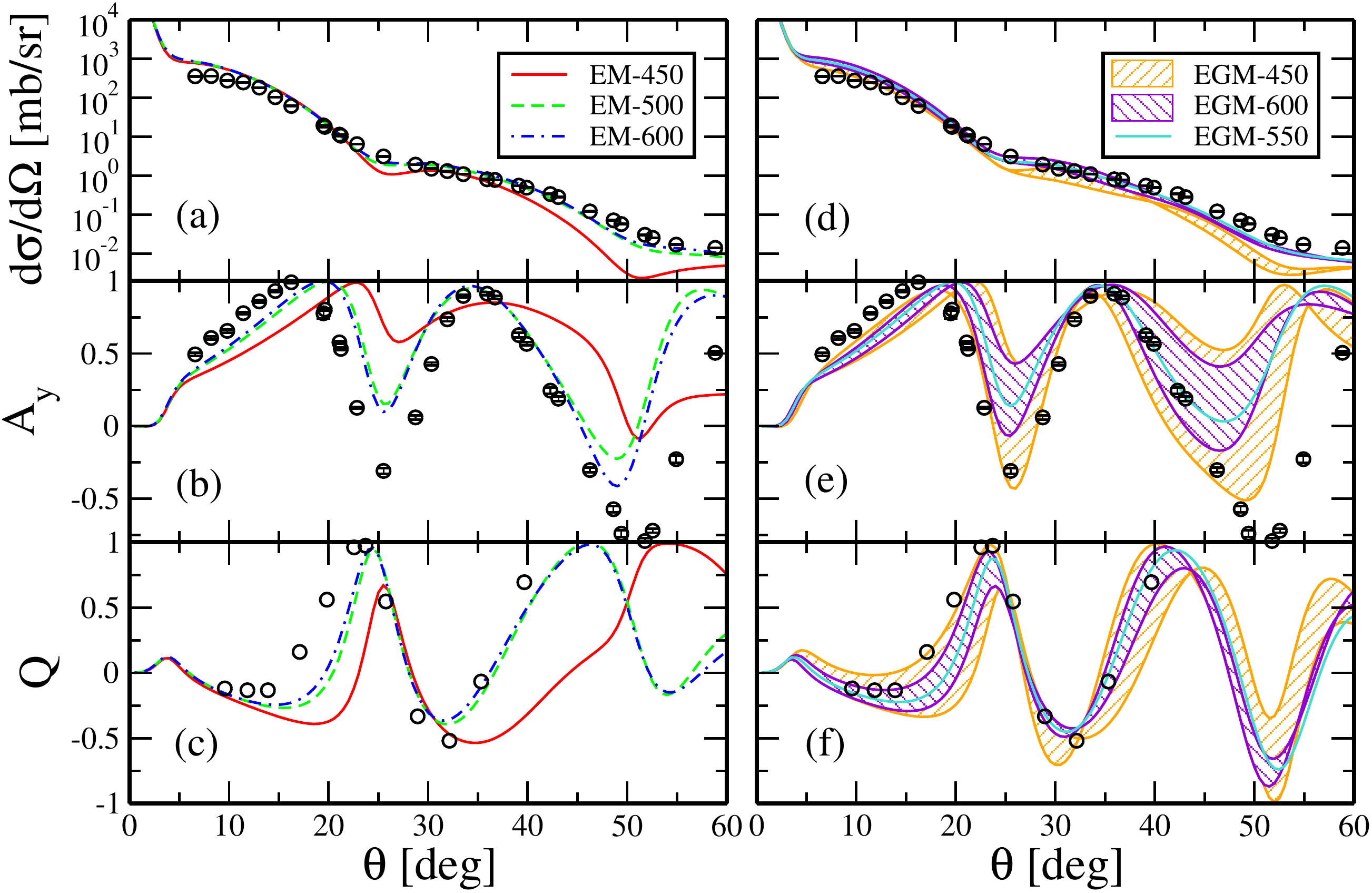}
\caption{\label{16O_200}  (Color online) The same as is Fig.~\ref{16O_100} but for an energy of $200$ MeV.
Data are taken from Refs.~\cite{kelly,exfor}.}
\end{center}
\end{figure}

\begin{figure}[t]
\begin{center}
\includegraphics[scale=0.33]{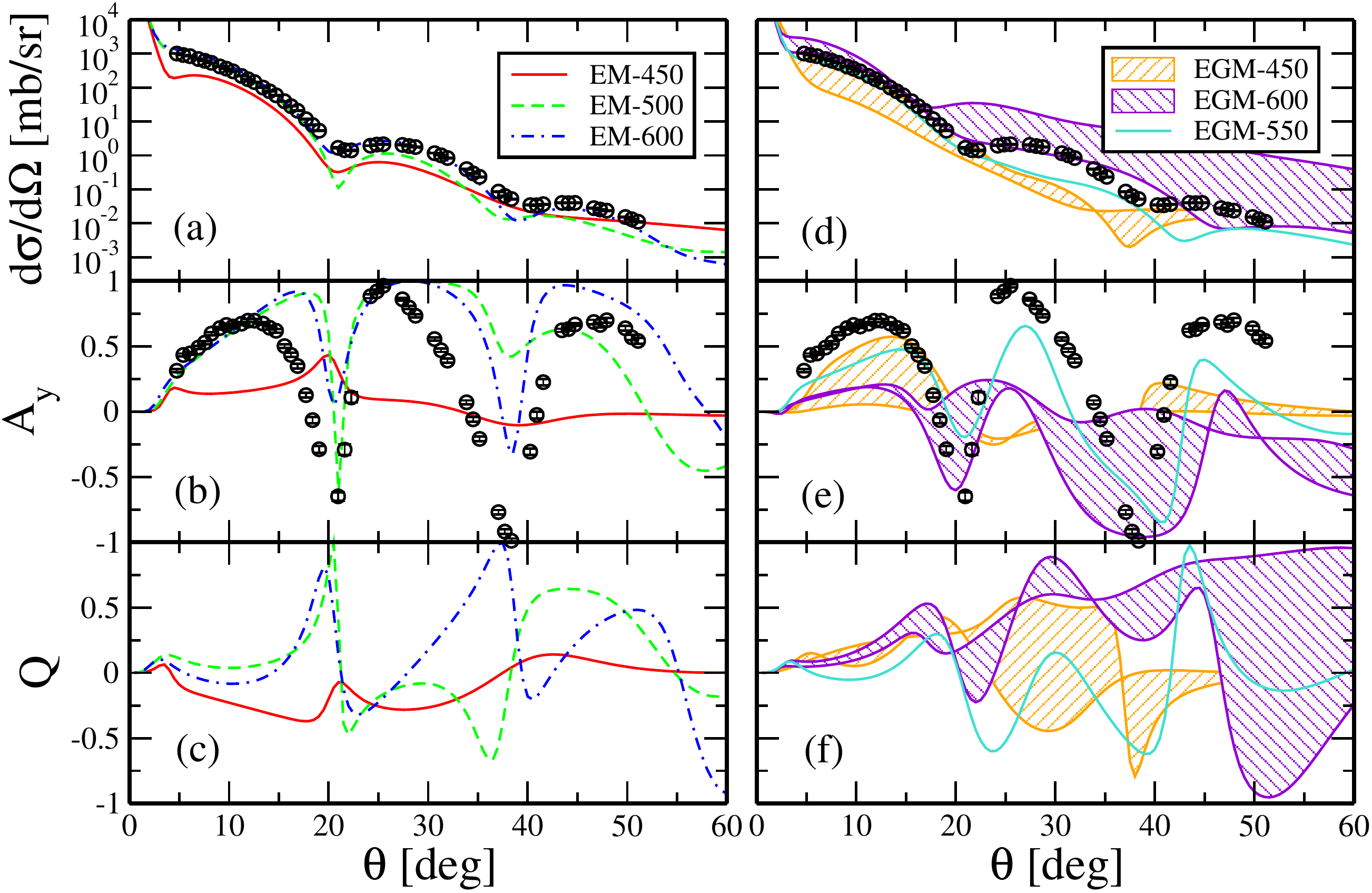}
\caption{\label{16O_318}  (Color online) The same as is Fig.~\ref{16O_100} but for an energy of $318$ MeV.
Data are taken from Refs.~\cite{kelly,exfor}.}
\end{center}
\end{figure}

\begin{figure}[t]
\begin{center}
\includegraphics[scale=0.33]{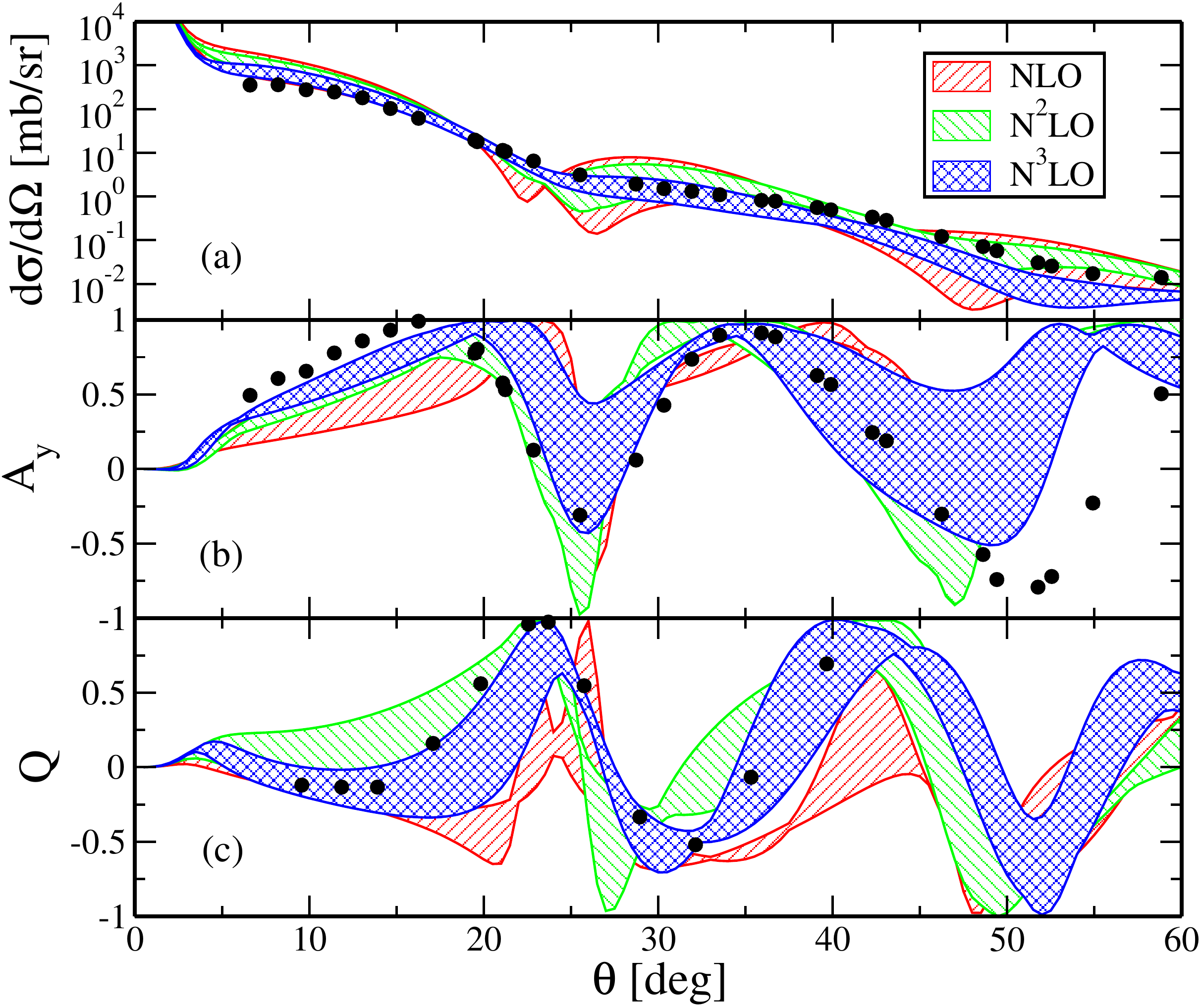}
\caption{\label{16O_orders} (Color online)
Scattering observables as a function of the center-of-mass scattering angle $\theta$  
for elastic proton scattering on ${}^{16}$O computed at $200$ MeV
(laboratory energy) with the EGM potential \cite{chiralepelbaum} at different orders: 
red bands are the NLO results, green and blue bands are respectively the \3bn2lo and \n3lo
results.  Data are taken from Refs.~\cite{kelly,exfor}.}
\end{center}
\end{figure}

\begin{figure}[t]
\begin{center}
\includegraphics[scale=0.33]{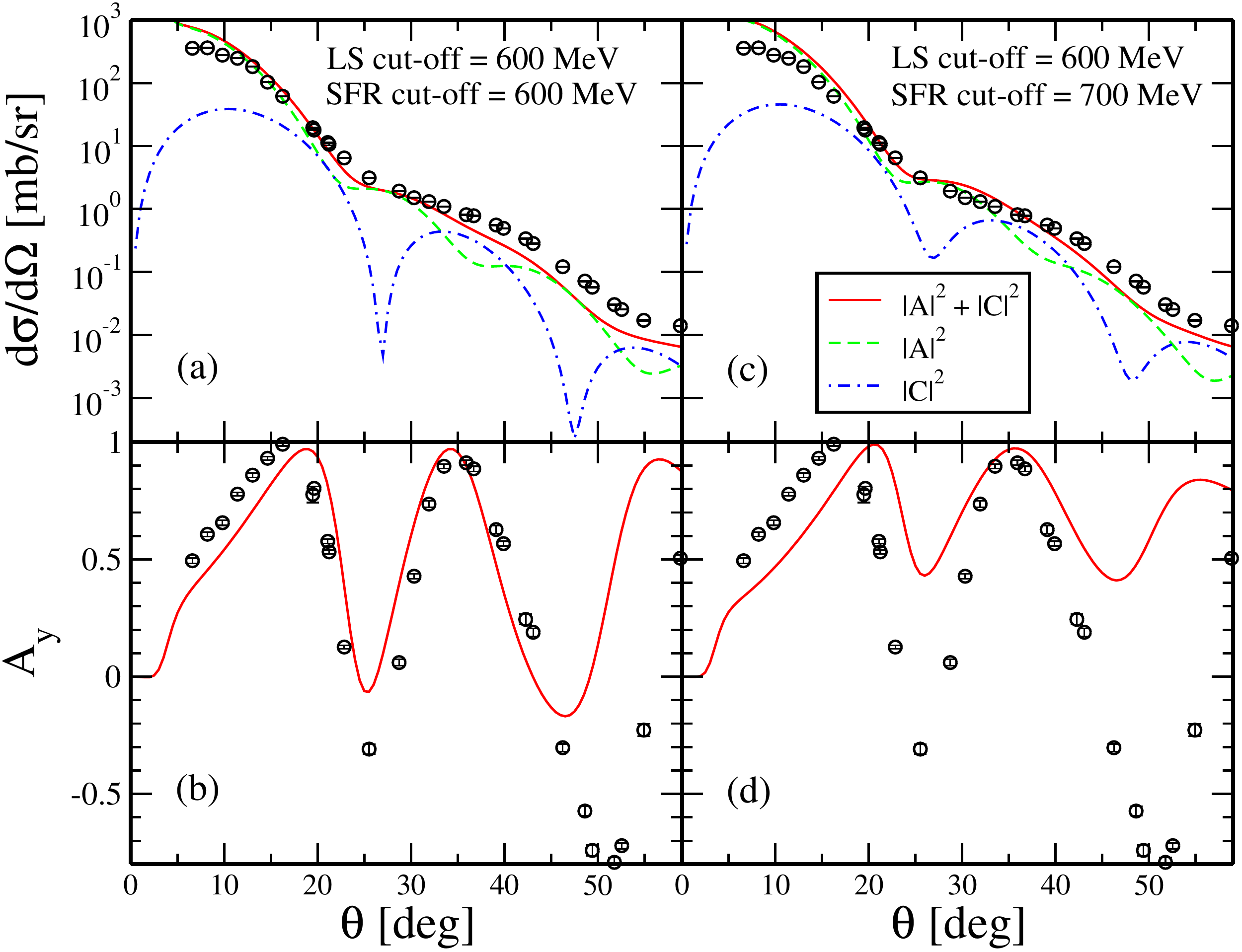}
\caption{\label{16O_contributions} (Color online)
Contributions, in terms of the amplitudes $A$ (\ref{naamplitudea}) and $C$ (\ref{naamplitudec}), 
to the scattering observables (differential cross section $d\sigma/d\Omega$ and
asymmetry parameter $A_y$) for elastic proton scattering on ${}^{16}$O computed at $200$ MeV
(laboratory energy) using two EGM potentials with 
$\{\Lambda, \tilde{\Lambda} \} = \{600,600 \},  \{600,700 \}$.
Coulomb distortion is included. 
Data are taken from Refs.~\cite{kelly,exfor}.}
\end{center}
\end{figure}

\begin{figure}[t]
\begin{center}
\includegraphics[scale=0.33]{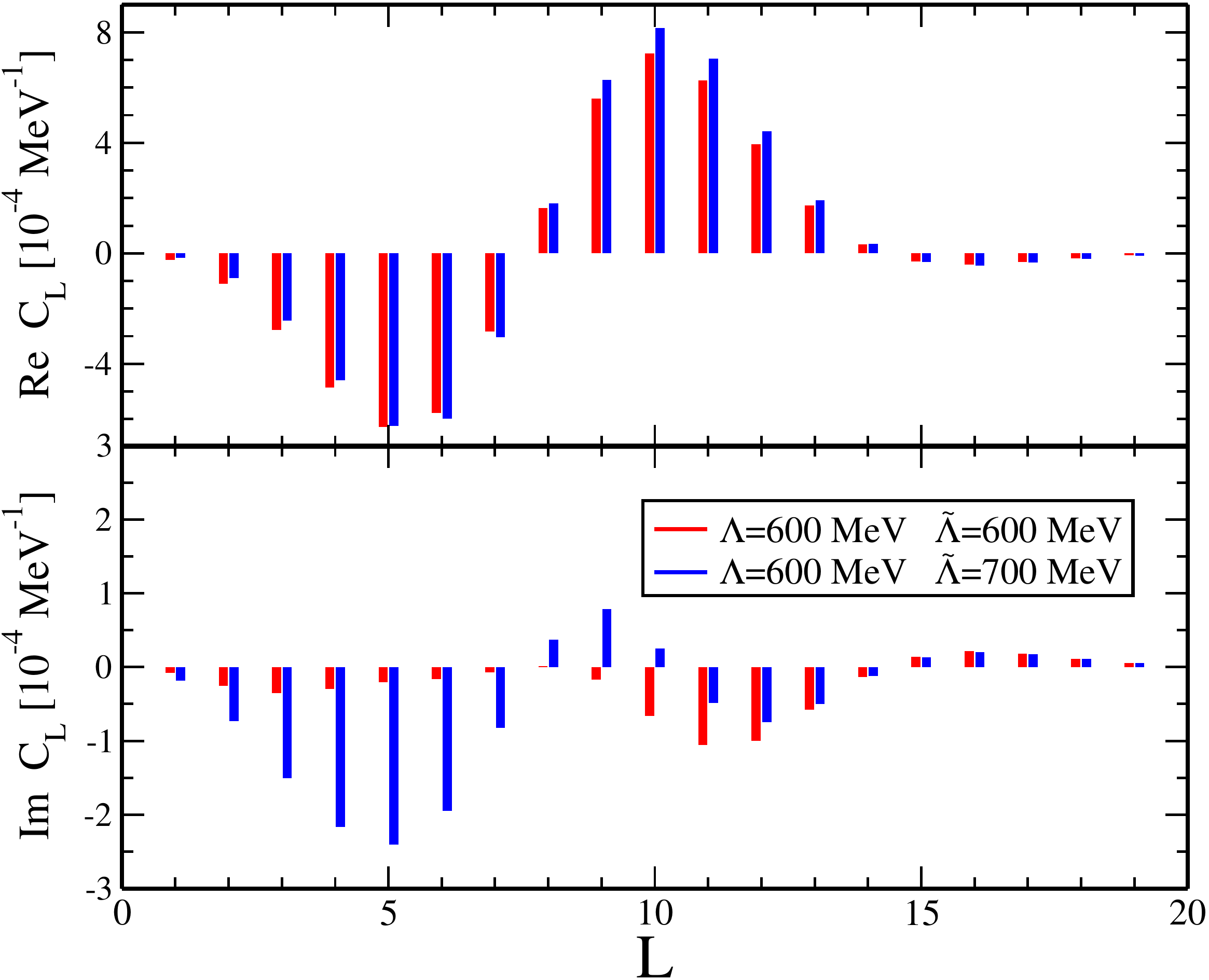}
\caption{\label{16O_components} (Color online) $L$-components (\ref{C_L}) for the $C$-amplitude for two EGM potentials with the following cutoffs: $\{\Lambda, \tilde{\Lambda} \} =
 \{600,600 \},  \{600,700 \}$.}
\end{center}
\end{figure}

In this section we present and discuss our numerical results for the $NA$ elastic scattering observables calculated with the microscopic
optical potential obtained within the theoretical framework described in Section~\ref{theofram}. As a study case in our calculations we
consider elastic proton scattering on $^{16}$O.

We investigate the sensitivity of our results to the choice of the $NN$ potential and, 
in particular, their dependence on the cutoff values. 
In order to investigate and emphasize the differences between the
different $NN$ potentials and also on the basis of the results obtained for the $NN$ Wolfenstein amplitudes $a$ and $c$, the scattering observables have been calculated for different energies ($100, 135, 200$, and $318$ MeV) for which experimental data are available. In light of the fact that chiral potentials are based upon a low-momentum expansion, the last energy may be considered beyond the limit of applicability of such potentials.

With these calculations we intend to achieve the following goals: 1) to check the agreement of our theoretical predictions with the empirical data; 2) to study the limits of applicability of chiral potentials in terms of the proton energy; 3) to identify the best set of values for the LS and, eventually, SFR cutoffs. 

In Figs.~\ref{16O_100}, \ref{16O_135}, \ref{16O_200}, and \ref{16O_318} we show the differential cross section ($d\sigma/d\Omega$), the analyzing power $A_y$, and the spin rotation $Q$ for elastic proton scattering on $^{16}$O as functions of the center-of-mass scattering angle $\theta$ with the above mentioned four energies ($E=100, 135, 200$, and $318$ MeV).
In the left panels we show the results obtained  with the EM 
potentials \cite{chiralmachleidt,PhysRevC.88.054002,PhysRevC.87.014322,PhysRevC.75.024311} 
while in the right panels we show the results obtained with the EGM potentials \cite{chiralepelbaum}. 
All potentials are denoted by the value of the LS cutoff.
The Coulomb interaction between the proton and the target nucleus is included as described in Sect. \ref{coulpot}.

In Fig.~\ref{16O_100}, at  $100$ MeV, all sets of potentials, regardless of cutoffs and theoretical approaches, give very similar results for all three observables, with the exception of $A_y$ above $50$ degrees, where
all potentials overestimate the experimental data up to the maximum and then display an unrealistic downward trend, and $Q$ around the maximum at $30$ degrees. 
In particular, the experimental cross section is well reproduced by all potentials in
the minimum region, between $30$ and $35$ degrees. Polarization observables are usually more sensitive to the differences in the potentials and to the ingredients and approximations of the model.
Experimental data for such observables are usually more difficult to reproduce.
Even if differences are rather small, potentials with the largest cutoff
($\Lambda = 600$ MeV) seem to provide the best description of $A_y$.

A similar result is obtained in Fig.~\ref{16O_135}, where we display the scattering observables calculated at $135$ MeV.
In this case all sets of potentials reproduce very well the experimental cross section and 
globally describe the shape of $A_y$, but are
unable to reproduce its magnitude for angles larger than $20$ degrees. 

In Fig.~\ref{16O_200} we plot the results obtained at $200$ MeV. At this energy, it is clear that potentials obtained with the lower cutoffs
(EM-450 and EGM-450) cannot be employed any further:  in both cases, the differential cross sections are not satisfactorily reproduced and the behaviour of $A_y$ and $Q$ as a function of $\theta$ is in clear disagreement with the empirical one. On the other hand, the remaining sets of potentials well describe the experimental cross sections and the analyzing power $A_y$, that is reasonably described not only for small scattering
angles but also for values larger than the minimum value up to about $45$ degrees. 

On the basis of all these results for $^{16}$O we can draw two conclusions: 
1) Potentials with lower cutoffs cannot reproduce experimental data at energies close to $200$ MeV.
2) There is no appreciable difference in using $500$ or $600$ MeV as LS cutoffs, even if the EM-$600$ and EGM-$600$ potentials seem to have a slightly better agreement with empirical data, in particular looking at polarization observables. 

For energies above $200$ MeV, this behaviour changes and the agreement with the experimental data begins to fail. 
This failure becomes larger as the energy increases.
As an example, in Fig.~\ref{16O_318} we display the results for the scattering observables on $^{16}$O computed at $318$ MeV, an energy for which experimental data are available.
We clearly see that at this energy all potentials are unable to describe the data. A somewhat better description is given by the EM-600 potential, which is able to reproduce  the global shape of the experimental results and the position of the minima, but the general agreement is poor. 
However, we stress that ChPT is a low-momentum expansion and its goal should be to perform calculations at lower energies.

In Fig.~\ref{16O_orders} we repeat the same order by order analysis (NLO, \3bn2lo, and \n3lo) 
of the convergence pattern performed in Sect. \ref{nnresults} for the NN amplitudes.
The results confirm the conclusion drawn looking at the NN amplitudes, 
i.e. that it is mandatory to use potentials at order \n3lo. At orders NLO and \3bn2lo our theoretical predictions not only underestimate or overestimate empirical data but also miss the overall shapes. The order by order convergence suggests that there is space for improvement going to higher orders (\chiral4lo) \cite{PhysRevLett.115.122301}.

In order to understand why some potentials provide a better description of certain
scattering data than other potentials, in Figs. \ref{16O_contributions} and \ref{16O_components} we plot the relevant components, Eqs. (\ref{naamplitudea}-\ref{naamplitudec}), for the differential cross section and the analyzing power computed at $200$ MeV. We have chosen, as a test case, two EGM potentials, with $\{\Lambda, \tilde{\Lambda} \} =
 \{600,600 \},  \{600,700 \}$, that reproduce differential cross sections with the same accuracy but give different predictions for the analyzing power.
In the upper panels of Fig. \ref{16O_contributions} we plot, for both potentials, the total differential cross section
(proportional to the sum $|A|^2 + |C|^2 $) with a red line, and the single contributions $|A|^2 $ and 
$|C|^2 $ with green and blue lines, respectively. The two potentials give similar results for $|A|^2 $ while significant differences around the minima are obtained for $|C|^2 $.
These differences, however, do not affect the final cross section, which is clearly dominated by the contribution proportional to $|A^2|$. The two potentials give relevant differences for the analyzing power $A_y$, which is plotted in the lower panels. In this case we cannot disentangle single contributions because $A_y$ is proportional to a combination of $A$ and $C$ ($A_y \sim \mathrm{Re} [A^{\ast} (\theta) \, C (\theta)]$). Nonetheless, a connection between $|C|^2 $ in the upper blue curves and $A_y$ seems to be plausible. 
To test if the first minimum of $A_y$ is really determined by the behaviour of the $C$ amplitudes, in Fig. \ref{16O_components} we plot the $L$-components of $C$, defined as 
\begin{equation}
\label{C_L}
C_L = \left[ F_L^+ (k_0) - F_L^- (k_0) \right] P_L^1 (\cos \theta) \; ,
\end{equation} 
evaluated at the angle $\theta=27^o$ corresponding to the minimum position.
The two potentials give close results for the real parts  and large differences  for the imaginary parts of the $L$-components.
With both potentials the real part of the $C$-amplitude is almost cancelled in the sum $C(\theta)= i/(2\pi^2) \sum_L C_L $.
For the imaginary part the sum gives a contribution that is small for $\{\Lambda, \tilde{\Lambda} \} = \{600,600 \}$ and sizable for 
$\{\Lambda, \tilde{\Lambda} \} = \{600,700 \}$.
As a consequence, in the case $\{\Lambda, \tilde{\Lambda} \} = \{600,600 \}$ the $C$-amplitude is very small and the analyzing power and the $|C|^2 $ contribution to the differential cross section develop well defined minima, while in the case $\{\Lambda, \tilde{\Lambda} \} = \{600,700 \}$, where the $C$-amplitude is larger, the corresponding minima are not deep enough and the disagreement with the experimental $A_y$  is more pronounced.

\section{Conclusions}
\label{concl}
In this work we have obtained a new microscopic optical potential for elastic proton-nucleus scattering. Our optical potential has been
derived as the first-order term within the spectator expansion of the nonrelativistic multiple scattering theory. 
In the interaction between the projectile and the target nucleon, which is described by the $NN$
$\tau$ matrix, we have neglected medium effects and we have adopted the impulse approximation, that consists in replacing $\tau$ by the
free $NN$ $t$ matrix.

As a further simplification, we have adopted the optimum factorization approximation, where the optical potential is given in a
factorized form by the product of the free $NN$ $t$ matrix and the nuclear density. This form conserves the off-shell nature of
the optical potential and it has been used in this work to compute the cross sections and the polarization observables of elastic
proton-nucleus scattering.

Two basic ingredients underlie the calculation of our microscopic optical potential: the $NN$ interaction and a model for nuclear
densities. For the $NN$ interaction we have used here for the first time the chiral potential. 
Microscopic optical potentials have been derived from two
different versions of the chiral potential at fourth order (\n3lo) based on the work of Entem 
and Machleidt (EM)~\cite{chiralmachleidt,PhysRevC.88.054002,PhysRevC.87.014322,PhysRevC.75.024311} 
and Epelbaum, Gl\"ockle, and Mei\ss ner (EGM)~\cite{chiralepelbaum}, which differ in the regularization scheme employed in the
two-pion exchange term and in the choice of the cutoffs. Neutron and proton densities have been obtained considering a system of
nucleons coupled to the exchange of mesons and the electromagnetic field through an effective Lagrangian. In practice, they have been
computed within the RMF description \cite{PhysRevC.66.024306} of spherical nuclei using a DDME model~\cite{PhysRevC.66.024306}. The Coulomb proton-nucleus
interaction has also been included in the calculations.

The $NN$ potentials have been used to calculate the $NN$ amplitudes that have then been employed to compute the $NN$ $t$ matrix. Results for
{\it pp} and {\it np} Wolfenstein amplitudes ($a$ and $c$) obtained with different $NN$ potentials have been presented and discussed.
Since ChPT is a low-momentum expansion of QCD, the agreement of the chiral potential with the experimental data becomes, as expected, worse increasing the energy. While at $100$ MeV all the $NN$ potentials are able to reproduce the experimental amplitudes, with the only exception of the real
part of $c_{pp}$ amplitude, that is anyhow extremely small, at $200$ MeV the set of potentials 
with lower cutoffs ($450$ MeV) fail to reproduce empirical data. 

As case study for our investigation we have considered elastic proton scattering on $^{16}$O. Results for the cross section, the
analyzing power, and the spin rotation have been presented and discussed in comparison with available experimental data.
Calculations have been performed with different $NN$ potentials at different energies.

The comparison
between the results obtained with the different versions of the chiral potential represents a useful test of the
reliability of our new optical potentials and allows us to identify the best set of LS cutoff values.

Polarization observables are more sensitive to the differences in the $NN$
interactions and to the approximations of the model. 
This sensitivity makes it difficult to describe the experimental analyzing powers
over the whole scattering angular distribution. 
The optical potentials obtained from all the $NN$ potentials give 
close results and a good description of the experimental cross
sections for proton energies up to about $135$ MeV. 
Of course, the differences among the results obtained with different $NN$ potentials increase with
the energy and with the scattering angle.
Our results indicate that EM-$600$  and EGM-$600$ provide a slightly better agreement with empirical data for energies up to $200$ MeV. Increasing the energy,
however, the agreement between the results from chiral potentials and data declines and it is plausible to
believe that above $200$ MeV ChPT is no longer applicable.

In the near future we plan to study the order by order convergence using the recent \chiral4lo potential \cite{PhysRevLett.115.122301} and to improve our 
calculations including three-body forces and nuclear medium effects. In addition, our investigation  will be extended to
$N \ne Z$ nuclei \cite{vorabbi}.

The case of elastic proton scattering considered in this work represents the first natural and necessary test of the reliability of an
optical potential. The optical potential, however, represents a crucial and critical input for calculations over a wide variety of
nuclear reactions and can therefore be employed in many other situations beyond those considered in this paper.


\section{Acknowledgements}
The authors are deeply 
grateful to E. Epelbaum (Institut f\"ur Theoretische Physik II,
Ruhr-Universit\"at Bochum) for providing the chiral potential of Ref. \cite{chiralepelbaum} and to R. Machleidt (Physics Department, University of Idaho, Moscow) for providing the chiral 
potentials of Refs. \cite{chiralmachleidt,PhysRevC.88.054002,PhysRevC.87.014322,PhysRevC.75.024311}

\bibliography{biblio}

\end{document}